\documentclass[10pt, journal]{IEEEtran}

\IEEEoverridecommandlockouts    

\usepackage{graphicx,lipsum}
\usepackage[export]{adjustbox}
\usepackage{mathtools, cuted}
\usepackage{booktabs}
\usepackage{comment}
\usepackage{authblk}
\usepackage{graphics}
\usepackage{threeparttable}
\usepackage[table,xcdraw]{xcolor}
\usepackage{multicol}
\usepackage{multirow}
\usepackage{tabularx}
\usepackage{xcolor}
\usepackage[linesnumbered,ruled,vlined]{algorithm2e}
\SetKwRepeat{Do}{do}{while}%
\usepackage{amsmath,amssymb}
\usepackage{caption}
\usepackage{paralist}
\usepackage{dblfloatfix}
\usepackage{subcaption}
\usepackage{setspace}
\usepackage{wrapfig}
\usepackage[hyphens]{url}
\usepackage{amsthm}
\usepackage{relsize}
\usepackage{amssymb}
\usepackage{pifont}
\usepackage{float}
\usepackage{graphicx}

\usepackage[table,xcdraw]{xcolor}

\makeatletter
\newcommand{\multiline}[1]{%
  \begin{tabularx}{\dimexpr\linewidth-\ALG@thistlm}[t]{@{}X@{}}
    #1
  \end{tabularx}
}
\makeatother

\makeatletter
\newcommand*\titleheader[1]{\gdef\@titleheader{#1}}
\AtBeginDocument{%
  \let\st@red@title\@title%
  \def\@title{%
    \bgroup\normalfont\large\centering\@titleheader\par\egroup
    \vskip1.5em\st@red@title}
}
\makeatother

\titleheader{© 2024 IEEE.  Personal use of this material is permitted.  Permission from IEEE must be obtained for all other uses, in any current or future media, including reprinting/republishing this material for advertising or promotional purposes, creating new collective works, for resale or redistribution to servers or lists, or reuse of any copyrighted component of this work in other works.
\\ 
\vspace{0.1in}The definitive version of this article has been accepted for publication in the IEEE Transactions on Information Forensics \& Security (TIFS)
}

\title{X-DFS: Explainable Artificial Intelligence Guided Design-for-Security Solution Space Exploration}

\author[1]{Tanzim Mahfuz}
\author[2]{Swarup Bhunia}
\author[1]{Prabuddha Chakraborty}
\affil[1]{Department of Electrical \& Computer Engineering, University of Maine, Orono, ME, USA}
\affil[2]{Department of Electrical \& Computer Engineering, University of Florida, Gainesville, FL, USA}

\date{}

\begin{document}

\maketitle

\thispagestyle{empty}
\pagestyle{plain}


\begin{abstract}
Design and manufacturing of integrated circuits predominantly use a globally distributed semiconductor supply chain involving diverse entities. The modern semiconductor supply chain has been designed to boost production efficiency, but is filled with major security concerns such as malicious modifications (hardware Trojans), reverse engineering (RE), and cloning. While being deployed, digital systems are also subject to a plethora of threats such as power, timing, and electromagnetic (EM) side channel attacks. Many Design-for-Security (DFS) solutions have been proposed to deal with these vulnerabilities, and such solutions (DFS) relays on strategic modifications (e.g., logic locking, side channel resilient masking, and dummy logic insertion) of the digital designs for ensuring a higher level of security. However, most of these DFS strategies lack robust formalism, are often not human-understandable, and require an extensive amount of human expert effort during their development/use. All of these factors make it difficult to keep up with the ever growing number of microelectronic vulnerabilities. In this work, we propose X-DFS, an explainable Artificial Intelligence (AI) guided DFS solution-space exploration approach that can dramatically cut down the mitigation strategy development/use time while enriching our understanding of the vulnerability by providing human-understandable decision rationale. We implement X-DFS and comprehensively evaluate it for reverse engineering threats (SAIL, SWEEP, and OMLA) and formalize a generalized mechanism for applying X-DFS to defend against other threats such as hardware Trojans, fault attacks, and side channel attacks for seamless future extensions.

\end{abstract}



\section{Introduction}
\begin{figure*}[]
\centering

\includegraphics[width=0.95\linewidth]{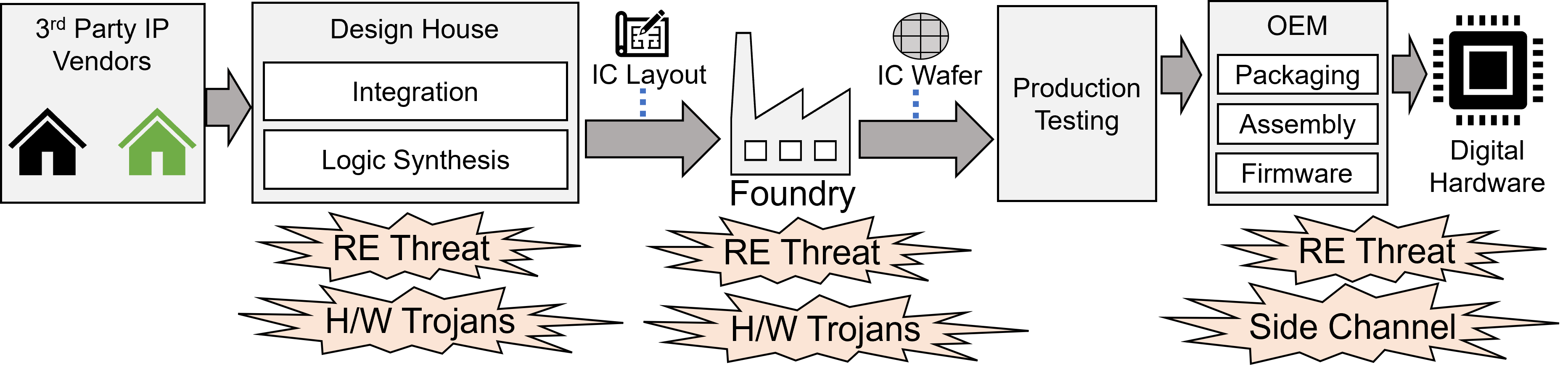}
\caption{Semiconductor supply chain security threats.\label{fig:supplychain}}

\vspace{0.1in}
\end{figure*}

A horizontal and distributed supply chain is at the heart of the booming semiconductor industry (see Fig.~\ref{fig:supplychain}). Creation of digital hardware such as integrated circuits (IC) typically involves the development of sub-designs (intellectual properties - IP) by small entities, the integration of 3rd party sub-designs with the in-house components at the main design house, IC layout creation, fabrication at a foundry, testing at a testing facility, and assembly by the original electronic manufacturer (OEM). This model reduces the time to market for digital systems, allows for specialization, and enables small businesses. All these steps are typically carried out by different entities in different geographical locations. Such a flow can lead to a series of problems, such as: (1) design or sub-design theft, (2) hardware cloning, and (3) malicious hardware modification/tampering. Microelectronic ICs and devices in the field also face diverse threats, such as power side channel attacks and reverse engineering.
A variety of  Design-for-Security (DFS) strategies such as IC metering \cite{Alkabani2007Active}, \cite{Koushanfar2012Provably}, watermarking \cite{kahng1998watermarking}, camouflaging \cite{Rajendran2013Is}, \cite{yasin2016camoperturb}, split manufacturing \cite{jarvis2007split}, \cite{imeson2013securing}, logic locking \cite{Roy2008EPIC:, sisejkovic2022logic}, gate parameter optimization \cite{karna}, variable delay module insertion \cite{PATCH}, and dummy logic insertion \cite{khaleghi2015fpga} have been developed to guarantee trust in the supply chain \cite{raj2023deepattack} and boost post-deployment IC/device security. However, security solutions often become obsolete with the emergence of novel attacks, while developing appropriate countermeasures requires extensive research effort, time, and expert resources. 

To expedite the solution search process (against novel vulnerabilities) and to create a human-understandable knowledge base of the said vulnerability, we propose an automated framework, X-DFS (E\textbf{X}plainable - \textbf{D}esign \textbf{F}or \textbf{S}ecurity). X-DFS uses a heuristics-based search process to determine a large set of DFS candidate instances that might contribute towards the defense against a given vulnerability. These DFS candidates are then used to train an explainable AI model that is capable of: (1) Emitting DFS rules that can be used to efficiently mitigate the vulnerability; (2) Automatically apply these rules towards securing the design. Hence X-DFS not only secures the design, it can also help researchers obtain a deeper understanding of the vulnerability. 

The proposed framework is highly generalized in nature and can be applied to a wide range of vulnerabilities (with minor tweaks), while most state-of-the-art DFS techniques are typically hand-crafted for mitigating a specific vulnerability or a small set of vulnerabilities. X-DFS can modify a design towards mitigating a given vulnerability and at the same time can generate human-understandable design transformation rules. Such capabilities do not exist in current state-of-the-art DFS techniques. X-DFS is highly flexible (parameterized) and extremely efficient in terms of computation cost, while most state-of-the-art DFS techniques are static in nature (not highly configurable) and fail to work for larger designs (inefficient).

\begin{figure}[!t]
\centering

\includegraphics[width=\linewidth]{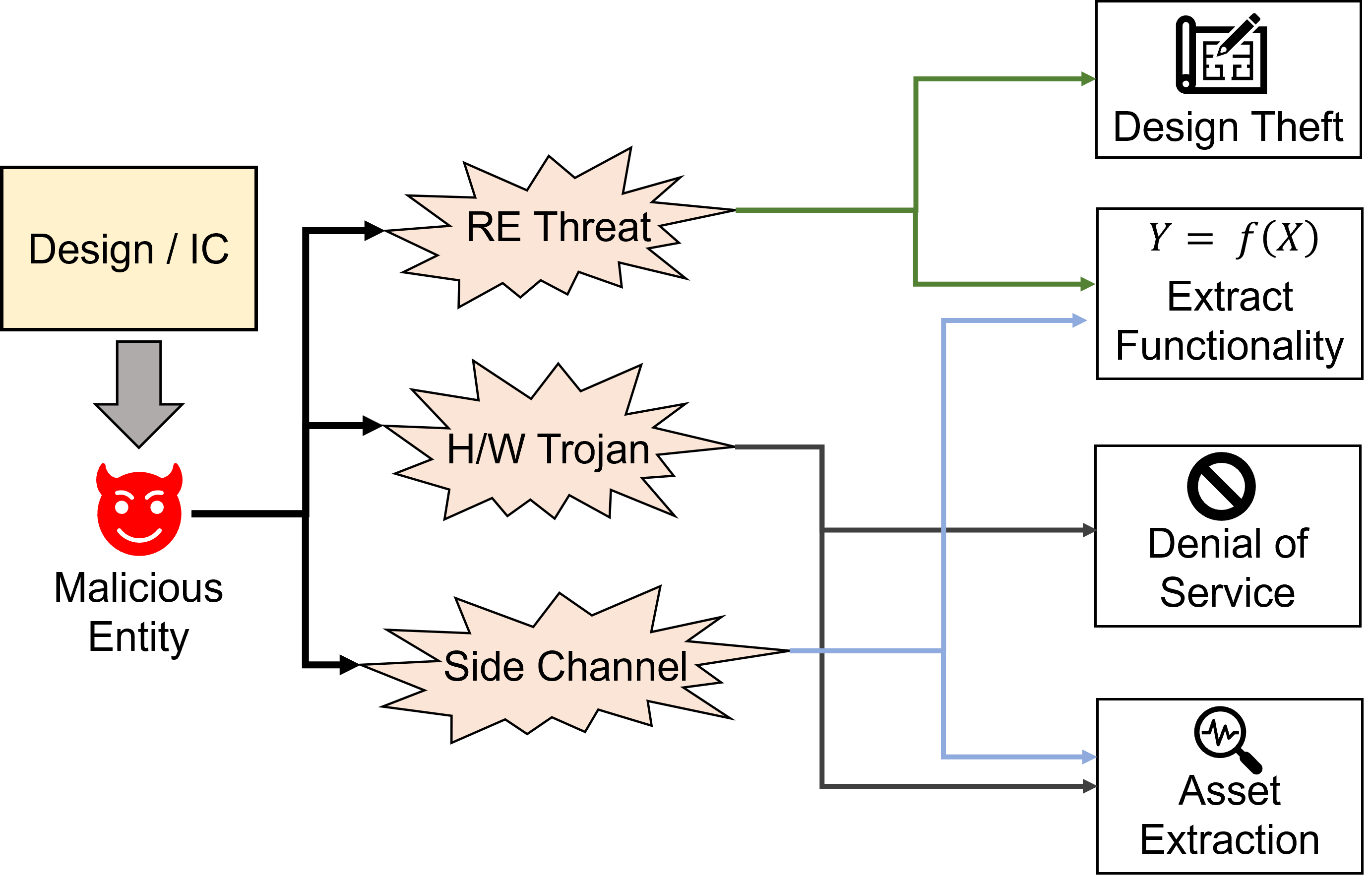}
\caption{Impacts of digital IC/design threats.\label{fig:RE}}

\vspace{-0.2in}
\end{figure}
\begin{figure}[!t]
\centering

\includegraphics[width=\linewidth]{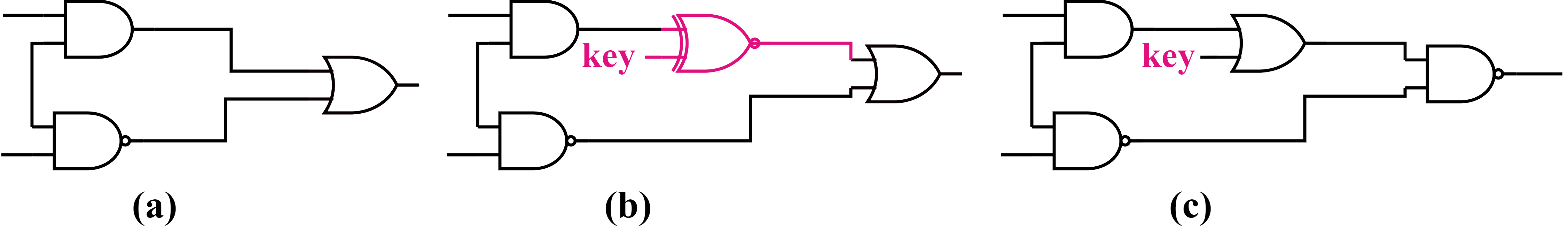}
\caption{Logic locking: (a) original netlist, (b) obfuscated netlist, and (c) synthesized netlist (structurally changed).\label{fig:LogicLocking}}

\vspace{-0.2in}
\end{figure}



We implement X-DFS as a highly parameterized robust framework and use it to perform a comprehensive effectiveness analysis for large-scale designs. We evaluate the X-DFS framework by testing it in the logic locking domain where X-DFS is used to automatically search for mitigation strategies (human-understandable) against three powerful logic locking attacks (SAIL \cite{chakraborty2021sail}, OMLA \cite{alrahis2021omla}), SWEEP \cite{alaql2019sweep}.
X-DFS was able to learn how to defend against these attack models and at the same time extracted human-understandable rules that can be used by other logic locking frameworks (such as LeGO \cite{alaql2021lego}) or human experts to carry out the locking process. 




In particular, we make the following research contributions:

\begin{enumerate}
    
    \item Formalize a general framework and the core mechanisms for automatic exploration of the design-for-security search space for countering novel attack vector(s).
    
    \item Design a set of algorithms (for reverse engineering attacks) that leverages this knowledge regarding an attack vector(s) to build an X-DFS model that can modify a given target design to be resilient against the attacks.

    \item Define a methodology to extract and understand the defense rules (human understandable) that are learned by the X-DFS models.

    \item Implement the proposed algorithms as a highly parameterized and scalable tool.

    \item Qualitatively and quantitatively verify the efficacy of the X-DFS framework/tool against different reverse engineering threats (SAIL, SWEEP, and OMLA).
    
\end{enumerate}


\section{Related Works \& Motivations}\label{sec:relWorks} 
Next, we provide the additional details about different DFS solutions and understand the motivations behind X-DFS. 

\subsection{Design for Security (DFS)}
Design for security techniques are widely used to protect digital designs and ICs from diverse attack vectors such as hardware Trojans, reverse engineering, and side channel threats (see Fig.~\ref{fig:RE}). These techniques typically involve the insertion of security elements in the design (like logic locking key gates) and strategic modification of existing logic. IC Metering techniques \cite{Alkabani2007Active} modifies the original finite state machine towards creating a Boosted Finite State Machine (BFSM) that ensures that the IC is only in the power-up state when an unique ID is provided. Watermarking techniques often insert an watermark in unused configurable logic blocks (CLBs) outputs \cite {kahng1998watermarking} to provide passive intellectual property theft protection. Techniques such as CamoPerturb \cite{yasin2016camoperturb} attempts to minimally perturb the design logic towards thwarting reverse engineering. Logic Locking (LL) involves inserting a set of additional gates into the design (connected to some key inputs) such that the modified design does not function correctly without the application of the right key values to these gates \cite{Amir2018Development, NEOS}. This ensures functional security of the design (to stop reverse engineering), but guessing the correct key input is almost trivial by investigating the structure of the gates inserted. For example, the correct key input for an additional XOR gate inserted to lock a wire will always be $0$. Hence, a set of structural changes are introduced by synthesizing the design to make key guessing difficult by observing the design structures (see Fig.~\ref{fig:LogicLocking}). Karna \cite{karna} attempts to mitigate power side channel leakage by: (1) analyzing each $N \times N$ grid of the design; (2) obtaining the corresponding TVLA-scores; (3) reconfiguring the gates if a vulnerability is detected in the grid. Variable delay module insertion has also been proposed to provide defense against side channel analysis \cite{PATCH}. To prevent the insertion of hardware Trojans in a design, researchers have proposed the insertion of dummy logic in the design \cite{khaleghi2015fpga}.

\subsection{Background on Logic Locking}
Several different types of logic locking techniques have been developed over the years. Combinational logic locking involves the insertion of combinational gates in the design towards corrupting the output if a wrong key value is applied \cite{Rajendran2015Fault,Baumgarten2010Preventing,Dupuis2014novel,Rajendran2012Security,Roy2008EPIC:}. Sequential locking techniques such as HARPOON \cite{Chakraborty2009HARPOON:} obfuscate the state space requiring a user to apply a sequence of correct key inputs towards unlocking the design. Trace logic locking (TLL) \cite{zuzak2020trace}, Delay-based locking \cite{xie2017delay}, Analog locking \cite{jayasankaran2020breaking}, and RTL locking techniques \cite{takhar2022holl} have also been developed to address different requirements. The main focus of this work is combinational logic locking, and let us start by looking at its brief history.

\begin{figure*}[!t]
\centering

\includegraphics[width=0.99\linewidth]{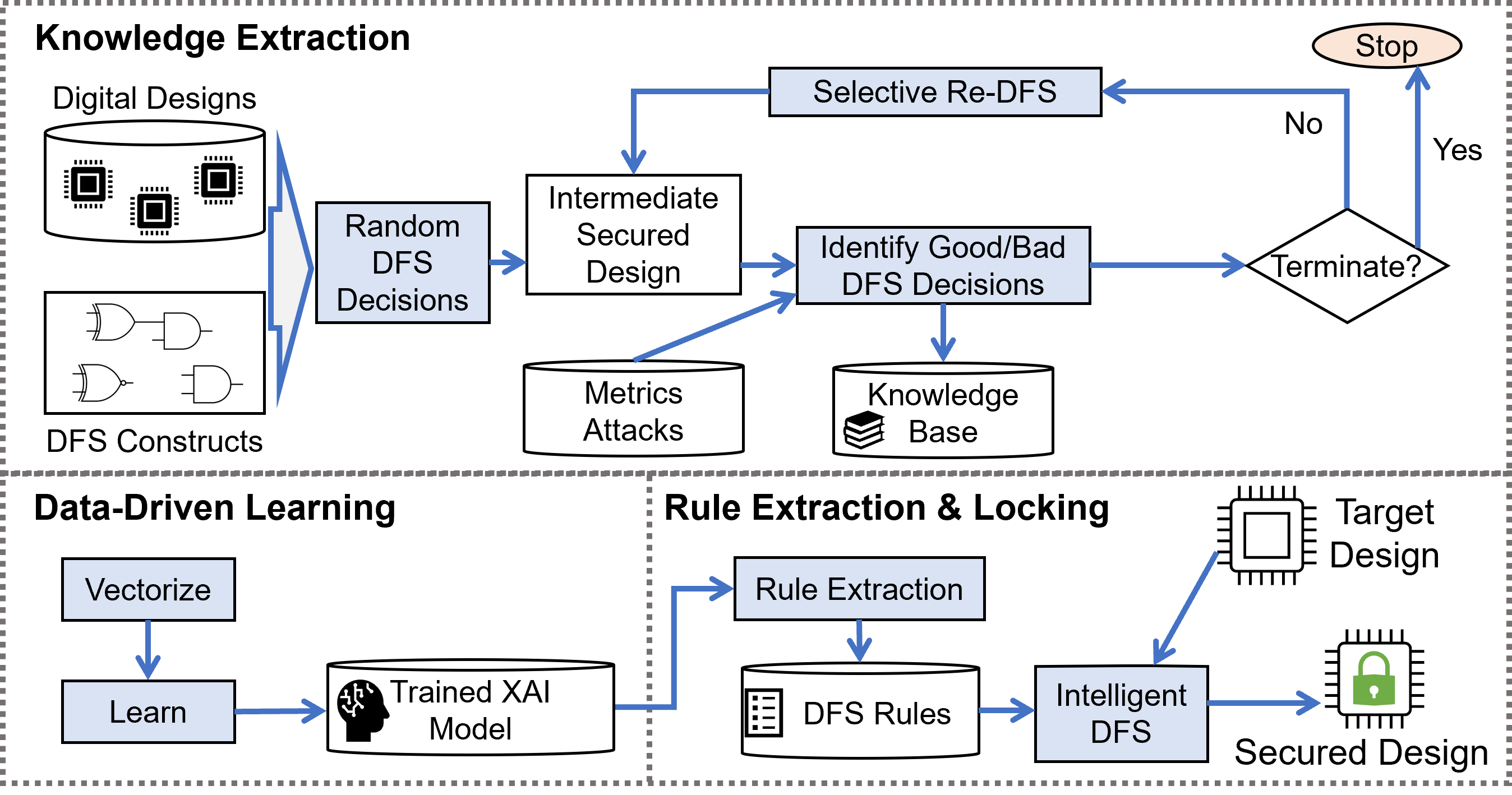}
\caption{Overview of the proposed X-DFS methodology.\label{fig:overview}}

\end{figure*}

\subsubsection{First Generation of LL}
The first set of techniques focused on inserting only XOR (key = 0) and XNOR (key = 1) gates randomly towards encrypting the digital design \cite{Amir2018Development, NEOS}. However, this led to suboptimal output corruption and unpredictable security against attacks such as KSA \cite{Yasin2016On} which leveraged several pitfalls in these locking schemes.
\begin{table*}[]
\centering
\captionsetup{justification=centering}
\caption{Notations and acronyms used in this article.}
\label{notations}
\renewcommand{\arraystretch}{1}
\small\addtolength{\tabcolsep}{6.2pt}

\begin{tabular}{|
>{\columncolor[HTML]{ECF4FF}}c |c|c|c|}
\hline
\cellcolor[HTML]{E2DDDD}\textbf{Notation} & \cellcolor[HTML]{E2DDDD}\textbf{Description} & \cellcolor[HTML]{E2DDDD}\textbf{Presented As} & \cellcolor[HTML]{E2DDDD}\textbf{Denoted in} \\ \hline
\textbf{X-DFS}      & Explainable Design For Security             & Framework & Entire Paper          \\ \hline
\textbf{XAI}        & Explainable AI                              & Concept  & Entire Paper           \\ \hline
\textbf{design or D}   & Gate Level Digital Design                              & String Variable  & Algorithm~\ref{algo:knowledge_extraction} and~\ref{algo:XAIL_Lock}, Section III-A \\ \hline
$\textbf{A}_\textbf{t}$   & Set of Attacks                              & Concept  & Section III-A \\ \hline
\textbf{C}   & Set of DFS Constructs                              & Concept  & Section III-A \\ \hline
\textbf{V}   & Vertices or Gates                              & Variable of Graph  & Section III-A and Section III-G \\ \hline
\textbf{E}   & Hyperedges  or Wires                              & Variable of Graph   & Section III-A and Section III-G \\ \hline
\textbf{keySize or KL} & Number of keys that needs to be inserted               & Integer Variable &   Algorithm~\ref{algo:knowledge_extraction} and~\ref{algo:XAIL_Lock}\\ \hline
\textbf{lockDict}   & Available logic gates to lock                    & String Variable & Algorithm~\ref{algo:knowledge_extraction} and~\ref{algo:XAIL_Lock}  \\ \hline
$\textbf{Th}_\textbf{it}$     & Number of iterations to terminate Algorithm~\ref{algo:knowledge_extraction}     & Integer Variable & Algorithm~\ref{algo:knowledge_extraction}   \\ \hline
\textbf{loc or lc}     & Size of locality                            & Integer Variable & Algorithm~\ref{algo:knowledge_extraction} and~\ref{algo:XAIL_Lock}    \\ \hline
$\textbf{Th}_\textbf{g}$        & Threshold to measure goodness               & Float Variable & Algorithm~\ref{algo:XAIL_Lock}   \\ \hline
\textbf{A}          & Flag to turn on Ada X-DFS                   & Boolean Variable  & Algorithm~\ref{algo:XAIL_Lock}    \\ \hline
\textbf{U}          & Flag to lock only distinct wires             & Boolean Variable & Algorithm~\ref{algo:XAIL_Lock}    \\ \hline
\textbf{RN}         & Flag to turn on randomness in locking       & Boolean Variable & Algorithm~\ref{algo:XAIL_Lock}    \\ \hline
\textbf{M}          & Multiplier to choose number of good locking & Integer Variable & Algorithm~\ref{algo:XAIL_Lock} \\ \hline
\end{tabular}%

\end{table*}
\subsubsection{Heuristics Driven Locking}
The next generation of locking schemes such as SLL \cite{Rajendran2012Security}, XOR/XNOR insertion based on fliprate \cite{NEOS}, and logic cone based insertion (CS) \cite{Amir2018Development} were designed to be more robust against vulnerabilities such as KSA \cite{Yasin2016On}. However, these heuristics-driven techniques have been shown to be vulnerable to SAT-based attacks that could extract the key by pruning the key space \cite{pramod2015, azar2019smt, Shamsi2017AppSAT:}. 

\subsubsection{Anti-SAT Locking Era}
A plethora of logic-locking techniques were developed to defend against attacks like SAT (and later SMT). These techniques either attempted to increase the time it took for SAT to complete each iteration or simply increased the number of iterations that SAT required to perform before it is able to obtain the key. Prominent among these techniques are: Anti-SAT, Strong Anti-SAT, SFLL, and CAS-Lock \cite{xie2018anti, liu2020strong, shakya2020cas, yasin2019sfll}. However, these techniques have vulnerabilities, including susceptibility to removal attacks and having low output corruption.  Full-Lock \cite{kamali2019full} is designed to exponentially increase the time required for each iteration of the SAT attack by making the underlying SAT problem harder to solve. It uses logarithmic self-routing networks to create SAT-hard instances by increasing the complexity of recursive calls. To enhance security, Full-Lock \cite{kamali2019full} introduces key-configurable inverters and replaces certain gates with small Spin Transfer Torque (STT) based Look-Up Tables (LUTs). Another novel work, LoPher \cite{saha2020lopher} is the first which combines hardware security with well-established cryptographic primitives like block ciphers. LoPher\cite{saha2020lopher} embeds combinational logic into a block cipher, using the S-Box and diffusion layers to perform gate-level operations.

\subsubsection{Rise of Structural Analysis}
SAIL \cite{Chakraborty2019SAIL:, chakraborty2021sail} was the first machine learning-guided structural analysis attack that was designed to identify the lack of structural obfuscation during logic locking. Due to poor structural security, one could carry out removal attacks (to even undermine SAT security), perform key guessing, and carry out reverse engineering. Following SAIL, a large range of structural analysis-based logic locking attacks emerged such as OMLA \cite{alrahis2021omla}, Snapshot \cite{sisejkovic2021challenging}, SCOPE \cite{alaql2021scope}, and SWEEP \cite{alaql2019sweep}. These attacks are highly scalable because they typically perform local analysis-based prediction, making their computations scale only with the size of key inputs (not the design size). Different defense techniques have been proposed to counter these attacks (e.g., UNSAIL \cite{unsail}).




\subsection{Why should we use XAI for DFS?} 
Most DFS methods follow two steps: (1) Determine where in the design modifications are required and (2) Determine the correct nature of these modifications. Assuming that there are $n$ valid regions in the design where modifications can be made for security and also let us assume that $m$ different types of modifications are possible. Then we have a total of $(m+1)^n$ possible design modification options (+1 to consider no-change). For example, if (1) we are performing logic locking on a design with $60,000$ gates; (2) we are allowed to insert XOR, XNOR, and AND gates; (3) we are to insert a 256 bit key, then we have a total of ${60,000 \choose 256} \times 3^{256}$ choices. Hence, the DFS search space is vast and navigating this requires expert crafted heuristics driven algorithms. Designing these algorithms can be time-consuming and will most likely require the involvement of domain experts. Hence, using an automated algorithm to the search the solution space can save significant amount of time and cost. DFS strategies get obsolete with the emergence of novel attacks requiring a tweak to the existing DFS strategy or the development of a new DFS strategy. Hence, an automated framework can be instrumental in rapidly dealing with these novel attacks. To build this automated framework, we propose to use Artificial Intelligence (AI) due to its wide spread success in automating many other tasks in diverse domains. Furthermore, we would like to incorporate explainability (hence, XAI) into the framework to enable the extraction of human-understandable DFS rules that are being learned by the AI model for dealing with a given attack(s).

\subsection{Related Works}
Isolated DFS solutions have been developed to mitigate reverse engineering \cite{Rajendran2012Security, NEOS, Amir2018Development, yasin2015improving}, side channel \cite{karna, PATCH}, and hardware Trojan \cite{khaleghi2015fpga} threats but there exists no systematic way to automatically devise defense strategies against novel attack vectors. Due to the rapidly growing microelectronic threat space, it has become extremely difficult for the research community and the industry to keep up with developing defense strategies against these novel attacks. These concerns are not addressed by current state-of-the-art DFS schemes. Moreover, techniques such as LeGO \cite{alaql2021lego} are designed to apply already known defense rules to mitigate threats and are not designed to generate new defense strategies. Our proposed framework, X-DFS promises to address these concerns making it distinct and highly impactful.


\section{X-DFS Methodology}\label{sec:method}
The overall framework is depicted in Fig.~\ref{fig:overview}. Next, we formally describe how X-DFS can be used to defend against different microelectronic threats and do a deeper dive into the methodology, particularly for reverse engineering threats.


\subsection{A Formal \& Generalized approach to X-DFS}
To formalize the X-DFS process, let us assume that:
\begin{enumerate}
    \item We have a digital design ($D = \{V, E\}$). $V$ = vertices/gates and $E$ = hyperedges among elements in $V$.
    \item There is a set of metrics/attacks ($A_t = \{a_1, a_2, ..., a_n\}$) that can evaluate the vulnerabilities of the design ($D$).
    \item There are a set of DFS constructs ($C = \{c_1, c_2, ..., c_k\}$) that might lead to the design being resilient to the said attacks ($A_t$) if inserted or utilized in the right way.
\end{enumerate}
With these assumptions, X-DFS first executes a knowledge extraction process by: (1) Randomly selecting $c_i$ from $C$; (2) Randomly selecting a location ($gate \in V$ or $wire \in E$) in $D$ to insert $c_i$; (3) Inserting $c_i$ and obtaining a modified design ($D'$); (4) Validating the effectiveness of such a modification ($D'$) by utilizing $A_t$ and tracking this information in the knowledge base ($K$); (5) Fully or partially reverting the design ($D'$) to its original state ($D$). After repeating steps (1-5) for a certain number of times we use the knowledge base ($K$) to craft potent DFS rules ($R$) that utilizes $C$ to defend designs beyond just $D$ against $A_t$.



\begin{table*}[]
\centering
\captionsetup{justification=centering}

\caption{X-DFS Generalization.}
\label{table:formalize}
\renewcommand{\arraystretch}{1}
\small\addtolength{\tabcolsep}{4pt}
$\textbf{A}_\textbf{t}$ 
\begin{tabular}{|
>{\columncolor[HTML]{ECF4FF}}c |c|c|}
\hline
\cellcolor[HTML]{EFEFEF}\textbf{} & \cellcolor[HTML]{EFEFEF}$\textbf{A}_\textbf{t}$ \textbf{(Attack, Metrics)} & \cellcolor[HTML]{EFEFEF}\textbf{C (DFS Constructs)} \\ \hline
\textbf{Reverse  Engineering}     & SAIL \cite{chakraborty2021sail}, OMLA \cite{alrahis2021omla}, SWEEP \cite{alaql2019sweep}                                    & Logic Locking Constructs  (XOR, XNOR, AND) \cite{Amir2018Development, NEOS}          \\ \hline
\textbf{Hardware Trojan}          & TRIT \cite{cruz2018automated}, MIMIC \cite{cruz2022machine}, TRIT-DS \cite{cruz2023framework}                               & Dummy Logic, Control Points \cite{khaleghi2015fpga}                        \\ \hline
\textbf{Side Channel}             & TVLA \cite{schneider2016leakage}, SVF \cite{kf2020param}                                           & Gate Resizing \cite{karna}, Masking Gates \cite{trichina2003combinational}                        \\ \hline
\textbf{Fault Analysis}           & SOLOMON \cite{srivastava2020solomon}, FaultDroid \cite{roy2020faultdroid}                                  & Speed-up,  Slow-down Constructs \cite{roy2022avatar}                     \\ \hline
\end{tabular}

\end{table*}

X-DFS can apply to the most microelectronic security threats if we can define $A_t$ and $C$ for that specific threat as shown in Table ~\ref{table:formalize}. The success of X-DFS and the layout of this approach depends on the the hypothesis: Randomly selecting elements in $C$ and inserting them in random locations of $D$ will lead to at least some positive outcomes in terms of defense against $A_t$. In this paper, we empirically validate this hypothesis using our reverse engineering case study results. Future works will attempt to extensively validate this hypothesis for other microelectronic threats. We also intend to augment this random search with heuristics from existing human-generated knowledge-base to speed up the search process. For example, for Trojans, we will attempt to focus our search around rarely trigger nodes since that is where Trojans are typically inserted.

\subsection{Understanding the Feature Set}
\label{sec:features}

\begin{figure}[!t]
\centering

\includegraphics[width=\linewidth]{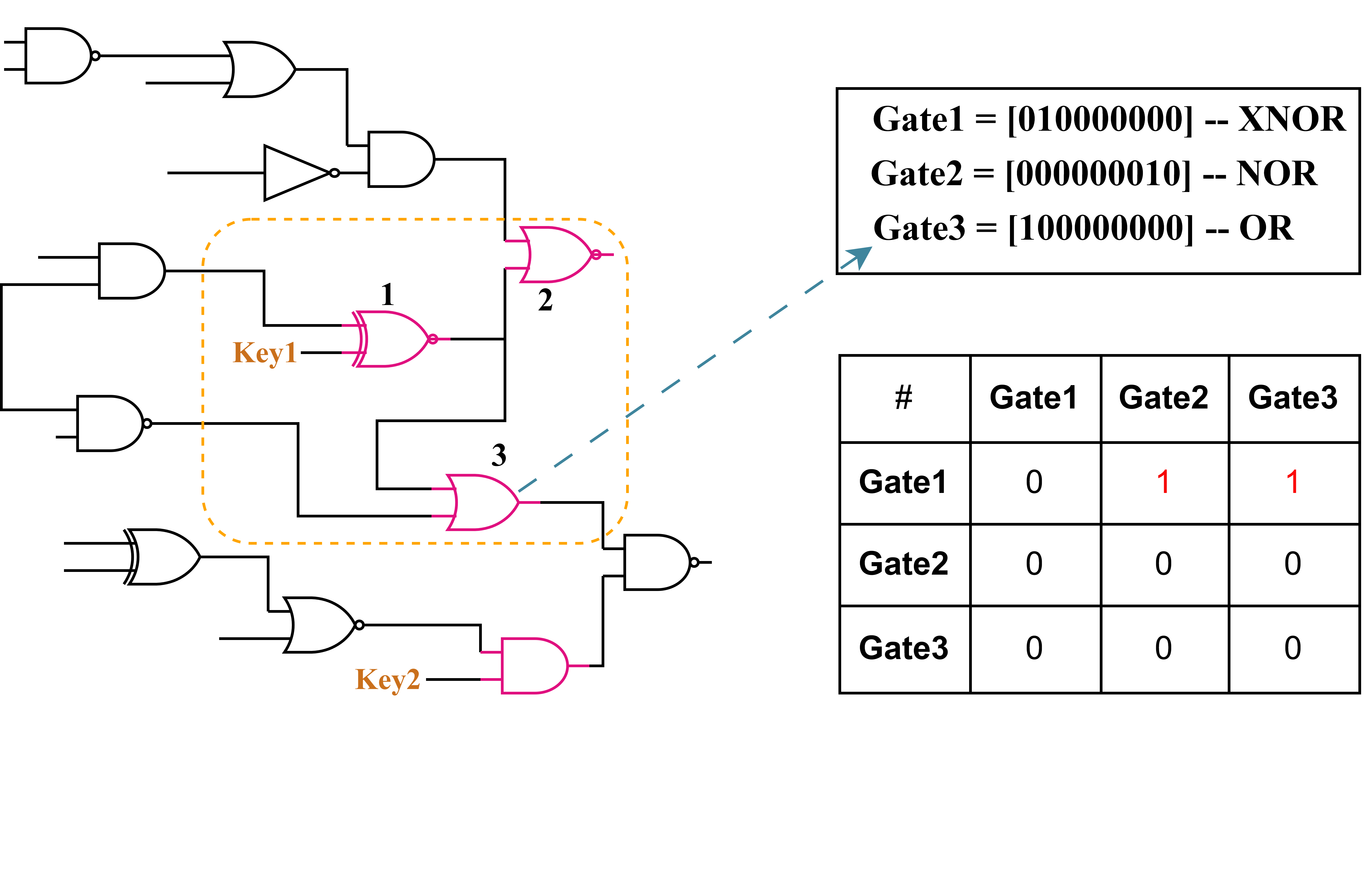}
\caption{Encoding a 3-gates sub-design structural features.\label{fig:structFeat}}

\vspace{-0.2in}
\end{figure}
Training and using an AI/XAI model will require us to first come up with a set of features that can capture diverse properties of a design or a sub-design (locality). For training and using X-DFS, we propose to use local structural features of design regions similar to SAIL \cite{chakraborty2021sail, Chakraborty2019SAIL:} and functional features (Static and Transition probabilities) of nearby wires based on methods proposed in \cite{gaikwadhardware}. The structural features of a wire captures information about the placement and connectivity of its locality (different types of gates and their interconnections). For a sub-design graph, we encode the connectivity as an adjacency matrix  and represent the gate types using a one-hot encoding scheme. The structural feature extraction process is depicted in Fig.~\ref{fig:structFeat} and more details can be found in the SAIL article \cite{chakraborty2021sail}. The gates are named ($G1$, $G2$, etc.) based on a breadth first search (BFS) algorithm search-order starting from the originating gate of the wire being considered for locking.

The functional features of a given wire are computed using the formulas presented in Table~\ref{table:static_feat} and Table~\ref{transproba} \cite{gaikwadhardware}. For a given signal or net in  Table~\ref{table:static_feat}, $P_A$ and $P_B$ are static probabilities (also known as signal probability) which indicate the fraction of time that the state is predicted to be at logic-1 or logic-0. If the net has a static probability of 0.4, then 40\% of the time it is anticipated to be at logic-1. In Table~\ref{transproba}, $A$ and $B$ represent the transition probability which is the number of jumps from logical level 0 to 1 or 1 to 0. The formulas in both of these tables are restricted to only the logic gates used to represent our benchmarks \cite{designs}. However, this list can be extended to support more complex gates (MUX, AOI, OAI) for future works. For computing functional features (during logic locking case study), the input wires are assumed to have a static and transition probability of $0.5$. 

Although we limit our current study to these specific features, the designed X-DFS framework can easily support additional structural and functional features of the design in the future. For example, it is feasible to incorporate features obtained from a graph neural network \cite{OMLA}, as well as other functional features such as fan-in, fan-out, clustering coefficient and centrality measurement \cite{hoque2018hardware_itc}.

\begin{table}[H]
\centering
\caption{Static probability computation formulas.}
\label{table:static_feat}
\renewcommand{\arraystretch}{}
\scriptsize\addtolength{\tabcolsep}{5pt}
\begin{tabular}{|c|c|}
\hline
 \textbf{Gate} & \textbf{Signal Probability of 1}  \\ \hline
 NOT                                         & 1 - $P_A$                                                                              \\ \hline
 AND                                         & $P_A * P_B$                                                                              \\ \hline
 OR                                         & $P_A + P_B - P_A * P_B$                                                                              \\ \hline

\end{tabular}
\end{table}
\begin{table}[H]
\centering
\caption{Transition probability computation formulas.}
\label{transproba}
\renewcommand{\arraystretch}{}
\scriptsize\addtolength{\tabcolsep}{5pt}
\begin{tabular}{|c|c|}
\hline
\textbf{Gate} & \textbf{$P_{0to1} = P_{out=0} * P_{out=1}$}  \\ \hline
 NOT                                         & 1 - $A * A$                                                                             \\ \hline
AND                                         & (1 - $A * B)*(A*B)$                                                                               \\ \hline
 OR                                         & (1 - $A$) $*$ (1 - $B$) $*$ (1 - (1 - $A$) $*$ (1 - $B$))                                                                             \\ \hline
NAND                                         & ($A * B$) $*$ (1 - $A * B$)                                                                              \\ \hline
NOR                                         & (1 - (1 - $A$) $*$ (1 - $B$)) $*$ (1 - $A$) $*$ (1 - $B$)                                                                               \\ \hline
XOR                                         & (1 - ($A$ + $B$ - 2 $*$ $A$ $*$ $B$)) $*$ ($A$ + $B$ - 2 $*$ $A * B$)                                                                           \\ \hline

\end{tabular}
\end{table}


\subsection{Automatic Knowledge Extraction}
As seen in Fig.~\ref{fig:overview}, we start with a set of reference digital designs, and a set of DFS constructs. In case of logic locking, these constructs are gates such as XOR, AND, etc. Next, we randomly insert a set of DFS constructs into the design, apply the selected attack vector, and identify which insertions were successfully attacked and which insertions remain secure. If the attack model bypasses/removes an inserted construct, we label this insertion (choice of both DFS construct + region of insertion) as 0, a bad label. If not, we label it 1, a good label. These labels and associated features are stored in the knowledge base. After that, we remove bad insertions, randomly add new DFS constructs, and continue the knowledge extraction process until a termination condition is met (based on runtime, iterations, or label ratio). The amount of randomly added new constructs will be equal to the number of removal of bad insertions. 

The essential consistent notations used in this paper are outlined in Table~\ref{notations}. We have thoroughly discussed those notations when necessary. Algo.~\ref{algo:knowledge_extraction}, provides greater details of the knowledge extraction process specifically for logic locking. Algo.~\ref{algo:knowledge_extraction} receives inputs such as the design ($design$) itself, possible locking gates ($lockDict$), number of keys to lock the design ($keySize$), locality size ($loc$) and number of iterations ($Th_{it}$) to terminate the algorithm. In line 1, we parse the input design file and capture the information using an internal graph data structure. In line 2, we pre-compute the functional features. In line 5, we initialize the $run$ variable that is used to track the number of iterations of the knowledge extraction process. In lines 6-23, we lock the design, perform the selected attack, extract the locking knowledge, relock the design, and repeat until a termination condition (no vulnerability is found or certain iterations by user's input are met, line 23) is reached. Inside this block, in line 7, we perform a random locking of the design by inserting an additional $keySize$ bits worth of key gates, where the locking constructs are also randomly sampled from the locking dictionary ($lockDict$). Next in line 8, we extract all the vulnerable and non-vulnerable key wires based on the $attack$ model. Next for each key wire region, we extract a set of structural (line 10) and functional (line 11) features. Next, the locking type associated with the specific key wire region is encoded as a vector (line 12 and line 13). If the vulnerability is found for that key insertion, then it is marked as $0$ (the attack is successful for this region), lines 14-15. Otherwise, we set the label as $1$ (considered a good locking against this attack), lines 16-17. The features \{S,F,L\} are inserted into $Train\_Data$ and the labels are inserted inside $Train\_Label$. The vulnerable key wires are removed (line 20) and the $keySize$ is reduced to the total number of vulnerable insertions ($count$) in line 21. In line 22, we increment the $run$ variable. We terminate this iterative knowledge extraction process when the $Vuln$ list is empty or when we have reached the user-specified maximum iteration threshold ($Th_{it}$). The dataset ($Train\_Data$ \& $Train\_Label$) is returned in line 24.

\begin{algorithm}[]
\DontPrintSemicolon 
\KwIn{$[lockDict, attack, design, keySize, loc, Th_{it}]$}
\KwOut{$[Train\_Data,Train\_Label]$}

$G \gets extract\_Graph(design)$ \;
$FnFeat \gets compute\_Functional\_Feat(G)$\;
$Train\_Data \gets []$ \;
$Train\_Label \gets []$ \;
$run \gets 0$ \;

\Do{$Vuln != empty$ and $run \le Th_{it}$}
{
    $G \gets random\_Lock(G, keySize, lockDict)$\;
    $Vuln, nonVuln \gets attack(G)$\;
    
    \For{$i$ \textbf{in} $(Vuln+nonVuln)$}
    {
        $S \gets ext\_Structural\_Feat(G, i, loc)$\;
        $F \gets FnFeat[i]$\;
        $lockType \gets find\_Type(G, i)$\;
        $L \gets encode\_Locking\_Type(lockType)$\;

        \If{$i$ in $Vuln$} 
        {
            $label \gets 0$\;
        }
        \Else
        {
             $label \gets 1$\;
        }

        $Train\_Data.append([S,F,L])$ \;
        $Train\_Label.append(label)$ \;
    }

    $[G, count] \gets remove\_Vuln\_Key(G, Vuln)$ \;
    $keySize \gets count$ \;
    $run++$ \;
    
}

\Return{$[Train\_Data,Train\_Label]$}\;
\caption{Knowledge Extraction}
\label{algo:knowledge_extraction}
\end{algorithm}



    



    


\begin{table}[]
\centering
\captionsetup{justification=centering}
\caption{SAIL accuracy for different X-DFS models. Algo.~\ref{algo:XAIL_Lock} parameters: $KL$ = 10\% of Design Size, $A=FALSE$, $RN=FALSE$, $U=TRUE$, $LC = 3$.}
\label{XAIL_AI_Models}
\renewcommand{\arraystretch}{1}
\small\addtolength{\tabcolsep}{6pt}
\begin{tabular}{|c|c|c|c|c|}
\hline
\rowcolor[HTML]{EFEDED} 
\multicolumn{1}{|c|}{\cellcolor[HTML]{E2DDDD}\textbf{Designs}}         & \multicolumn{1}{c|}{\cellcolor[HTML]{E2DDDD}\textbf{DT}} & \multicolumn{1}{c|}{\cellcolor[HTML]{E2DDDD}\textbf{ENSM}} & \multicolumn{1}{c|}{\cellcolor[HTML]{E2DDDD}\textbf{RF}} & \multicolumn{1}{c|}{\cellcolor[HTML]{E2DDDD}\textbf{ADB}} 
\\ \hline
\multicolumn{1}{|c|}{\cellcolor[HTML]{ECF4FF}\textbf{sqrt}}      & \multicolumn{1}{c|}{40.83}                                 & \multicolumn{1}{c|}{22.09}                                  & \multicolumn{1}{c|}{17.41}                                & \multicolumn{1}{c|}{2.10}                                                              
\\ \hline
\multicolumn{1}{|c|}{\cellcolor[HTML]{ECF4FF}\textbf{sin}}      & \multicolumn{1}{c|}{52.32}                               & \multicolumn{1}{c|}{15.40}                                  & \multicolumn{1}{c|}{12.20}                               & \multicolumn{1}{c|}{1.47}                                                     
\\ \hline
\multicolumn{1}{|c|}{\cellcolor[HTML]{ECF4FF}\textbf{div}}      & \multicolumn{1}{c|}{36.3}                              & \multicolumn{1}{c|}{15.72}                                 & \multicolumn{1}{c|}{29.79}                                & \multicolumn{1}{c|}{20.14}                                                                     
\\ \hline
\multicolumn{1}{|c|}{\cellcolor[HTML]{ECF4FF}\textbf{arbiter}}     & \multicolumn{1}{c|}{33.53}                                & \multicolumn{1}{c|}{6.93}                                  & \multicolumn{1}{c|}{11.13}                               & \multicolumn{1}{c|}{0.24}                                                              
\\ \hline
\multicolumn{1}{|c|}{\cellcolor[HTML]{ECF4FF}\textbf{memctrl}}      & \multicolumn{1}{c|}{47.14}                                & \multicolumn{1}{c|}{26.3}                                 & \multicolumn{1}{c|}{28.17}                             & \multicolumn{1}{c|}{13.03}                                                                  
\\ \hline
\multicolumn{1}{|c|}{\cellcolor[HTML]{ECF4FF}\textbf{log2}}      & \multicolumn{1}{c|}{42.83}                                    & \multicolumn{1}{c|}{13.33}                                & \multicolumn{1}{c|}{12.72}                                    & \multicolumn{1}{c|}{0.74}                                                                                
\\ \hline
\multicolumn{1}{|c|}{\cellcolor[HTML]{ECF4FF}\textbf{square}}      & \multicolumn{1}{c|}{33.31}                                & \multicolumn{1}{c|}{20.93}                                 & \multicolumn{1}{c|}{20.79}                                & \multicolumn{1}{c|}{13.56}                                                                               
\\ \hline
\multicolumn{1}{|c|}{\cellcolor[HTML]{ECF4FF}\textbf{multiplier}}      & \multicolumn{1}{c|}{40.13}                                & \multicolumn{1}{c|}{19.97}                                 & \multicolumn{1}{c|}{12.77}                                & \multicolumn{1}{c|}{0.69}                                                             
\\ \hline
\multicolumn{1}{|c|}{\cellcolor[HTML]{ECF4FF}\textbf{voter}}      & \multicolumn{1}{c|}{61.70}                                & \multicolumn{1}{c|}{30.78}                                 & \multicolumn{1}{c|}{28.39}                                & \multicolumn{1}{c|}{24.77}                                                       
\\ \hline

\multicolumn{1}{|c|}{\cellcolor[HTML]{ECF4FF}\textbf{Average}}                            & \multicolumn{1}{c|}{\textbf{43.12}}                               & \multicolumn{1}{c|}{\textbf{19.05}}                                  & \multicolumn{1}{c|}{\textbf{19.20}}                                & \multicolumn{1}{c|}{\textbf{8.52}}                                          
\\ \hline

\end{tabular}
\end{table}

\begin{table*}[]
\captionsetup{justification=centering}
\caption{Comparison of X-DFS with other prominent Logic Locking algorithms. Values indicate SAIL attack accuracy. Algo.~\ref{algo:XAIL_Lock} parameters: $KL=128$, $A=FALSE$, $RN=FALSE$, $U=TRUE$, $LC = 3$.}
\label{otherLocking}
\renewcommand{\arraystretch}{1}
\small\addtolength{\tabcolsep}{4pt}
\begin{tabular}{|c|c|c|c|c|c|c|}
\hline
\rowcolor[HTML]{EFEDED} 
\multicolumn{1}{|c|}{\cellcolor[HTML]{E2DDDD}\textbf{Designs}}         & \multicolumn{1}{c|}{\cellcolor[HTML]{E2DDDD}\textbf{Random}} & \multicolumn{1}{c|}{\cellcolor[HTML]{E2DDDD}\textbf{SFLL\_Point \cite{NEOS,sfll}}} & \multicolumn{1}{c|}{\cellcolor[HTML]{E2DDDD}\textbf{Anti-SAT} \cite{NEOS}, \cite{xie2018anti}} & \multicolumn{1}{c|}{\cellcolor[HTML]{E2DDDD}\textbf{XORProb \cite{NEOS}}} & \multicolumn{1}{c|}{\cellcolor[HTML]{E2DDDD}\textbf{AOR \cite{NEOS}}} & \multicolumn{1}{c|}{\cellcolor[HTML]{E2DDDD}\textbf{X-DFS [ADB]}}
\\ \hline
\multicolumn{1}{|c|}{\cellcolor[HTML]{ECF4FF}\textbf{sqrt}}      & \multicolumn{1}{c|}{54.69}                                 & \multicolumn{1}{c|}{42.96}                                  & \multicolumn{1}{c|}{46.09}                                & \multicolumn{1}{c|}{56.25}                                     & \multicolumn{1}{c|}{50.78}                                 & \multicolumn{1}{c|}{0}                                 
\\ \hline
\multicolumn{1}{|c|}{\cellcolor[HTML]{ECF4FF}\textbf{sin}}      & \multicolumn{1}{c|}{53.12}                               & \multicolumn{1}{c|}{52.34}                                  & \multicolumn{1}{c|}{(-)}                               & \multicolumn{1}{c|}{47.65}                                & \multicolumn{1}{c|}{55.46}                                & \multicolumn{1}{c|}{4.69}                                  
\\ \hline
\multicolumn{1}{|c|}{\cellcolor[HTML]{ECF4FF}\textbf{div}}      & \multicolumn{1}{c|}{46.88}                              & \multicolumn{1}{c|}{50}                                 & \multicolumn{1}{c|}{32.03}                                & \multicolumn{1}{c|}{44.53}                                 & \multicolumn{1}{c|}{53.13}                                & \multicolumn{1}{c|}{0}                                               
\\ \hline
\multicolumn{1}{|c|}{\cellcolor[HTML]{ECF4FF}\textbf{arbiter}}     & \multicolumn{1}{c|}{53.90}                                & \multicolumn{1}{c|}{48.44}                                  & \multicolumn{1}{c|}{43.75}                               & \multicolumn{1}{c|}{42.19}                                  & \multicolumn{1}{c|}{57.03}                                & \multicolumn{1}{c|}{0}                                               
\\ \hline
\multicolumn{1}{|c|}{\cellcolor[HTML]{ECF4FF}\textbf{memctrl}}      & \multicolumn{1}{c|}{62.50}                                & \multicolumn{1}{c|}{55.47}                                 & \multicolumn{1}{c|}{44.53}                             & \multicolumn{1}{c|}{48.43}                                  & \multicolumn{1}{c|}{63.28}                                & \multicolumn{1}{c|}{3.13}                                             
\\ \hline
\multicolumn{1}{|c|}{\cellcolor[HTML]{ECF4FF}\textbf{log2}}      & \multicolumn{1}{c|}{49.21}                                    & \multicolumn{1}{c|}{44.53}                                & \multicolumn{1}{c|}{(-)}                                    & \multicolumn{1}{c|}{51.56}                                  & \multicolumn{1}{c|}{50}                                    & \multicolumn{1}{c|}{0}                                                      
\\ \hline
\multicolumn{1}{|c|}{\cellcolor[HTML]{ECF4FF}\textbf{square}}      & \multicolumn{1}{c|}{49.21}                                & \multicolumn{1}{c|}{50.78}                                 & \multicolumn{1}{c|}{(-)}                                & \multicolumn{1}{c|}{42.96}                                & \multicolumn{1}{c|}{52.34}                                & \multicolumn{1}{c|}{7.03}                                                  
\\ \hline
\multicolumn{1}{|c|}{\cellcolor[HTML]{ECF4FF}\textbf{multiplier}}      & \multicolumn{1}{c|}{51.56}                                & \multicolumn{1}{c|}{45.31}                                 & \multicolumn{1}{c|}{56.25}                                & \multicolumn{1}{c|}{48.43}                                & \multicolumn{1}{c|}{47.65}                                & \multicolumn{1}{c|}{1.56}                                                   
\\ \hline
\multicolumn{1}{|c|}{\cellcolor[HTML]{ECF4FF}\textbf{voter}}      & \multicolumn{1}{c|}{53.13}                                & \multicolumn{1}{c|}{46.88}                                 & \multicolumn{1}{c|}{55.46}                                & \multicolumn{1}{c|}{53.90}                                & \multicolumn{1}{c|}{54.68}                                & \multicolumn{1}{c|}{0}                                                 
\\ \hline

\multicolumn{1}{|c|}{\cellcolor[HTML]{ECF4FF}\textbf{Average}}                            & \multicolumn{1}{c|}{\textbf{52.69}}                               & \multicolumn{1}{c|}{\textbf{48.52}}                                  & \multicolumn{1}{c|}{\textbf{46.35}}                                & \multicolumn{1}{c|}{\textbf{48.43}}                                  & \multicolumn{1}{c|}{\textbf{53.81}}                                & \multicolumn{1}{c|}{\textbf{1.82}}                                         
\\ \hline

\end{tabular}
\end{table*}
\begin{table}[]
\centering
\captionsetup{justification=centering}
\caption{X-DFS against OMLA attack. Values indicate OMLA attack accuracy. Algo.~\ref{algo:XAIL_Lock} parameters: $KL=64$, $A=FALSE$, $RN=FALSE$, $U=TRUE$, $LC = 3$.}
\label{omla_Result}
\renewcommand{\arraystretch}{1}
\small\addtolength{\tabcolsep}{14pt}
\begin{tabular}{|c|c|c|}
\hline
\rowcolor[HTML]{EFEDED} 
\multicolumn{1}{|c|}{\cellcolor[HTML]{E2DDDD}\textbf{Designs}}         & \multicolumn{1}{c|}{\cellcolor[HTML]{E2DDDD}\textbf{Random}} & \multicolumn{1}{c|}{\cellcolor[HTML]{E2DDDD}\textbf{X-DFS [RF]}}  
\\ \hline
\multicolumn{1}{|c|}{\cellcolor[HTML]{ECF4FF}\textbf{sqrt}}      & \multicolumn{1}{c|}{95.31}                                                                   & \multicolumn{1}{c|}{9.37}                                                                        
\\ \hline
\multicolumn{1}{|c|}{\cellcolor[HTML]{ECF4FF}\textbf{sin}}      & \multicolumn{1}{c|}{67.19}                                                                & \multicolumn{1}{c|}{9.37}                                                             
\\ \hline
\multicolumn{1}{|c|}{\cellcolor[HTML]{ECF4FF}\textbf{div}}      & \multicolumn{1}{c|}{67.19}                                                             & \multicolumn{1}{c|}{4.68}                                                                                   
\\ \hline
\multicolumn{1}{|c|}{\cellcolor[HTML]{ECF4FF}\textbf{arbiter}}     & \multicolumn{1}{c|}{70.31}                                                                 & \multicolumn{1}{c|}{43.75}                                                                                    
\\ \hline
\multicolumn{1}{|c|}{\cellcolor[HTML]{ECF4FF}\textbf{memctrl}}      & \multicolumn{1}{c|}{95.31}                                                                 & \multicolumn{1}{c|}{48.44}                                                                                        
\\ \hline
\multicolumn{1}{|c|}{\cellcolor[HTML]{ECF4FF}\textbf{log2}}      & \multicolumn{1}{c|}{93.75}                                                                    & \multicolumn{1}{c|}{23.44}                                                                                                                   
\\ \hline
\multicolumn{1}{|c|}{\cellcolor[HTML]{ECF4FF}\textbf{square}}      & \multicolumn{1}{c|}{78.13}                                                                 & \multicolumn{1}{c|}{4.68}                                                                                                             
\\ \hline
\multicolumn{1}{|c|}{\cellcolor[HTML]{ECF4FF}\textbf{multiplier}}      & \multicolumn{1}{c|}{89.06}                                       & \multicolumn{1}{c|}{6.25}                                                                                      
\\ \hline
\multicolumn{1}{|c|}{\cellcolor[HTML]{ECF4FF}\textbf{voter}}      & \multicolumn{1}{c|}{70.31}                                                             & \multicolumn{1}{c|}{23.44}                                                                              
\\ \hline

\multicolumn{1}{|c|}{\cellcolor[HTML]{ECF4FF}\textbf{Average}}                            & \multicolumn{1}{c|}{\textbf{80.73}}                                                              & \multicolumn{1}{c|}{\textbf{19.27}}                                            
\\ \hline

\end{tabular}
\end{table}

\begin{table}[]
\centering
\captionsetup{justification=centering}
\caption{X-DFS against SWEEP. Values indicate SWEEP accuracy. Algo.~\ref{algo:XAIL_Lock} parameters: $KL$ = 10\% of Design Size, $A=FALSE$, $RN=FALSE$, $U=TRUE$, $LC = 3$.}
\label{table:sweep_result}
\renewcommand{\arraystretch}{1}
\small\addtolength{\tabcolsep}{14pt}
\begin{tabular}{|c|c|c|}
\hline
\rowcolor[HTML]{EFEDED} 
\multicolumn{1}{|c|}{\cellcolor[HTML]{E2DDDD}\textbf{Designs}}         & \multicolumn{1}{c|}{\cellcolor[HTML]{E2DDDD}\textbf{Random}} & \multicolumn{1}{c|}{\cellcolor[HTML]{E2DDDD}\textbf{X-DFS [RF]}}  
\\ \hline
\multicolumn{1}{|c|}{\cellcolor[HTML]{ECF4FF}\textbf{sqrt}}      & \multicolumn{1}{c|}{72.95}                                                                   & \multicolumn{1}{c|}{15.73}                                                                        
\\ \hline
\multicolumn{1}{|c|}{\cellcolor[HTML]{ECF4FF}\textbf{sin}}      & \multicolumn{1}{c|}{77.45}                                                                & \multicolumn{1}{c|}{6.74}                                                             
\\ \hline
\multicolumn{1}{|c|}{\cellcolor[HTML]{ECF4FF}\textbf{div}}      & \multicolumn{1}{c|}{79.95}                                                             & \multicolumn{1}{c|}{8.83}                                                                                   
\\ \hline
\multicolumn{1}{|c|}{\cellcolor[HTML]{ECF4FF}\textbf{arbiter}}     & \multicolumn{1}{c|}{69.87}                                                                 & \multicolumn{1}{c|}{28.36}                                                                                    
\\ \hline
\multicolumn{1}{|c|}{\cellcolor[HTML]{ECF4FF}\textbf{memctrl}}      & \multicolumn{1}{c|}{75.59}                                                                 & \multicolumn{1}{c|}{18.77}                                                                                        
\\ \hline
\multicolumn{1}{|c|}{\cellcolor[HTML]{ECF4FF}\textbf{log2}}      & \multicolumn{1}{c|}{78.75}                                                                    & \multicolumn{1}{c|}{25.08}                                                                                                                   
\\ \hline
\multicolumn{1}{|c|}{\cellcolor[HTML]{ECF4FF}\textbf{square}}      & \multicolumn{1}{c|}{85.83}                                                                 & \multicolumn{1}{c|}{10.85}                                                                                                             
\\ \hline
\multicolumn{1}{|c|}{\cellcolor[HTML]{ECF4FF}\textbf{multiplier}}      & \multicolumn{1}{c|}{73.19}                                       & \multicolumn{1}{c|}{28.32}                                                                                      
\\ \hline
\multicolumn{1}{|c|}{\cellcolor[HTML]{ECF4FF}\textbf{voter}}      & \multicolumn{1}{c|}{82.60}                                                             & \multicolumn{1}{c|}{15.89}                                                                              
\\ \hline

\multicolumn{1}{|c|}{\cellcolor[HTML]{ECF4FF}\textbf{Average}}                            & \multicolumn{1}{c|}{\textbf{77.35}}                                                              & \multicolumn{1}{c|}{\textbf{17.62}}                                            
\\ \hline

\end{tabular}
\end{table}

\subsection{Building the X-DFS Model}
\label{sec:ModelTraning}
Next, we train an XAI model to capture the DFS experimentation knowledge base.
All the $Train\_Data$ and $Train\_Label$ obtained from each design using Algo.~\ref{algo:knowledge_extraction} are merged to form $Train\_Data\_Merged$ and $Train\_Label\_Merged$. This dataset is used to fit a wide range of AI models that are either inherently explainable (such as Decision Tree) or can be explained using other techniques such as SHAP (SHapley Additive exPlanations) \cite{SHAP}, LIME (Local Interpretable Model-agnostic Explanations) \cite{LIME}, and Anchors \cite{anchors:aaai18}. We also experiment with different data preprocessing techniques (e.g., standard scalar, min/max scaling \cite{scikit-learn}). Although the Decision Tree (DT) and Random Forest (RF) model are by default explainable, we also obtain results with Ensembling (Decision Tree, Random Forest, and Adaboost). When considering the ensemble results, we assign equal importance to each model. Due to the random nature of the development of the dataset, if we encounter any data imbalance, we employ the SMOTE \cite{chawla2002smote} technique to address the imbalance. SMOTE, unlike basic oversampling methods that simply replicate instances from the minority class, produces new synthetic examples. This technique mitigates overfitting by increasing diversity and addressing class imbalance in classification problems. For all machine learning models, we utilize the default settings (as specified in \cite{scikit-learn}) except for the Adaboost learning rate (set to $0.01$).






\begin{algorithm}[]
\DontPrintSemicolon 
\KwIn{$[D, KL, Model, lockDict, A, U, RN, Th_{g}, lc, M]$}
\KwOut{$D\_Locked$}
$G \gets extract\_Graph(D)$ \;
$FnFeat \gets compute\_Functional\_Feat(G)$\;
$options \gets []$\;
$Good \gets []$\;
$breakFlag \gets FALSE$\;
\For{$i \gets 0$ \textbf{to} $len(G.wires)$} 
{
    \For{$j \gets 0$ \textbf{to} $len(lockDict)$} 
    {
        
        $S \gets ext\_Structural\_Feat(G, G.wires[i], lc)$\;
        $F \gets FnFeat[G.wires[i]]$\;
        $L \gets encode\_Locking\_Type(lockDict[j])$\;
         $P \gets Model([S,F,L])$\;
        
        $options.append([G.wires[i], lockDict[j], P])$\;
        \If{$A == TRUE$} 
        {
            $Good \gets find\_Good(options, Th_{g})$\;
            \If{$len(Good) \ge M \times KL$} 
            {
                $breakFlag \gets TRUE$\;
                $break$ \; 
            }
                           
        }
        
    }
    \If{$breakFlag == TRUE$}
    {
        $break$ \;
    }
}
\If{$A == TRUE$ and $RN == FALSE$} 
{
    $Good \gets Good.sort\_Descend\_ByP()$\;
}
\If{$A == TRUE$ and $RN == TRUE$} 
{
    $Good \gets Good.sort\_Descend\_ByP()$\;
    $Good \gets Good.shuffle()$\;
}
\If{$RN == TRUE$ and $A == FALSE$} 
{
    $Good \gets find\_Good(options, Th_{g})$\;
    $Good \gets Good.shuffle()$\;
}

\If{$RN == FALSE$ and $A == FALSE$} 
{
    $Good \gets options.sort\_Descend\_ByP()$\;
}
\If{$U == TRUE$} 
{
    $Good \gets unique\_Wire(Good)$\;
}

$G \gets lock(G, Good[0:KL])$ \;
$D\_Locked \gets write\_To\_File(G)$

\Return{$D\_Locked$}\;
\caption{X-DFS\_Lock}
\label{algo:XAIL_Lock}
\end{algorithm}

\subsection{Extraction of Design Transformation Rules}
We leverage different explainability algorithms towards visualizing the internal learning and decision making process of X-DFS models. Complex models perform best; however, they are not inherently human-understandable. Hence, we utilize the SHAP (SHapley Additive exPlanations) algorithms, specifically the Kernel Explainer (model-agnostic) towards understanding the feature importance \cite{SHAP}. SHAP uses a game theory methodology to determine which features contribute the most towards making the final prediction. For a given prediction, the Shapley value of a feature is its expected (weighted by probability) marginal contribution. The Shapley value of a feature $i$, $\phi_i$ is computed as shown in equation~\ref{eqn_shapley}. Here we assume that there are $p$ features and hence $p!$ is the total number of ways to form a coalition of $p$ features (players). Assume that $g$ is a given coalition. Then $\frac{|g|!(p-|g|-1)!}{p!}$ is the weight term and $val(g\cup\{i\})-val(g)$ is the marginal contribution of the feature $i$ to a given coalition $g$.

\begin{equation}
\label{eqn_shapley}
\phi_i = \sum_{g\subseteq\{1,...,p\}\{i\}}^{} \frac{|g|!(p-|g|-1)!}{p!} [val(g\cup\{i\})-val(g)]
\end{equation}

This Shapley value equation is derived based on four axioms: (1) Efficiency, (2) Symmetry, (3) Null Player, and (4) Additivity. The additivity axiom conveys that: if we combine two games (predictions), then the sum of a feature's contribution is its total contribution (assuming each prediction is independent). For more details, refer to \cite{shapley1951notes, SHAP}. For an X-DFS model, understanding the importance of each feature for different samples will allow us to understand what the X-DFS model has actually learned and distil that understanding into human-understandable DFS rules.  



\subsection{DFS using the XAI Model}
With a trained $X-DFS\_Model$, we can also directly use it (and the inherent DFS rules) to transform a target design. This process is described in Algo.~\ref{algo:XAIL_Lock} specifically for logic locking. Algo.~\ref{algo:XAIL_Lock} requires the design ($D$), number of key length ($KL$), the X-DFS model ($Model$), possible gates to lock ($lockDict$), participation of Ada X-DFS ($A$), participation of unique wires ($U$), participation of RN X-DFS ($RN$), threshold to determine the goodness of locking ($Th_g$), locality size ($lc$), integer value to filter Ada X-DFS ($M$) as inputs. In line 1, we parse the input design ($D$) and capture it in our graph data structure as $G$. In line 2, we pre-compute the functional features of every wire. Then for each wires in $G$ and for each entry in the locking dictionary $lockDict$, we obtain a real-valued prediction (0 to 1) using $X-DFS\_Model$ (lines 6-19). Next the specific design wire ($G.wires[i]$), the locking construct ($lockDict[j]$), and the prediction ($P$) are appended to $options$ (line 12). To speed up the locking process (and avoid unnecessary computation), we terminate this exhaustive evaluation process only when the $A$ (X-DFS\_Ada) flag is set to `TRUE' (line 13) and at least $M \times KL$ good locking options are discovered (lines 14-17). Here $M$ is a multiplier integer value provided as input by the user and $KL$ is the target key length for the locking operation. The goodness of a locking option is determined using a threshold ($Th_{g}$) imposed on the prediction values. 
Based on the goodness threshold ($Th_g$), the best $options$ are filtered into $Good$ in line 14. If $A$ (X-DFS\_Ada) flag is set to `TRUE' and $RN$ (X-DFS\_RN) is `FALSE', we use prediction value in descending order to sort the $Good$ candidates in lines 20-21.  To allow flexibility and some variability (to divert attackers), we also introduce some controlled randomness if $RN$ (X-DFS\_RN) is set to `TRUE' and if $A$ is set to `FALSE'. 
This is achieved by finding the top options ($Good$) using $Th_{g}$ (line 26), and then shuffling the list in line 27. To introduce randomness in the X-DFS\_Ada both $A$ and $RN$ need to be `TRUE' (shown in line 22-24). If both $RN$ and $A$ are set to `FALSE', then we simply utilize prediction value to sort the $options$ in descending order to create $Good$ (keeping only the top locking options) in lines 28-29. Locking can also be done for distinct wires by setting $U$ is `TRUE' (unique nets) in line 30-31. Finally, the design is locked with the top $KL$ elements in $Good$ and the resulting design graph is converted to the design format ($D\_Locked$) before being returned.

\subsection{Algorithms Complexity Analysis}
For Algo.~\ref{algo:knowledge_extraction}, let us assume that the termination condition (line 23) is based on a set number of iterations ($I$), there are $V$ gates in the design, there are $E$ wires in the design, and the time complexity of the attack/metric is $\mathcal{O}(A_t)$ (varies). Then the computational time complexity of line 1 is $\mathcal{O}(V + E)$, line 2 is $\mathcal{O}(V + E)$, line 7 (for each outer iteration) is $\mathcal{O}(KeySize)$, line 8 (for each outer iteration) is $\mathcal{O}(A_t)$, line 9-19 (for each outer iteration) is $\mathcal{O}(KeySize)$, line 20 (for each outer iteration) is $\mathcal{O}(KeySize)$, and line 21-22 (for each outer iteration) is $\mathcal{O}(Const)$. Then the total computational time complexity considering the outer loop, line 23 ($I$) is: $\mathcal{O}(V + E + V + E + I * KeySize + I * A_t * Const + I * KeySize + I * KeySize + I)$ = $\mathcal{O}(V + E + I * KeySize + I * A_t)$. Both $I$ and $KeySize$ will be small values and will not necessarily scale across applications. Hence, the complexity can be further simplified to $\mathcal{O}(V + E + A_t)$. For attacks such as SAIL where the computation required is $\mathcal{O}(KeySize)$, the time complexity becomes $\mathcal{O}(V + E)$ (considering $KeySize$ is small).

For Algo.~\ref{algo:XAIL_Lock}, let us also assume that the locking dictionary ($lockDict$) size is $L$. Then the computational time complexity of line 1 is $\mathcal{O}(V + E)$, line 2 is $\mathcal{O}(V + E)$, lines 8-12 (for each iteration) is $\mathcal{O}(Const)$ where $Const$ is constant, line 14 (for each iteration) is $\mathcal{O}(EL)$, line 21 is $\mathcal{O}(EL*Log(EL))$, lines 23-24 is $\mathcal{O}(EL*Log(EL) + EL)$, lines 26-27 is $\mathcal{O}(EL)$, line 29 is $\mathcal{O}(EL*Log(EL))$, line 31 is $\mathcal{O}(EL)$, line 32 is $\mathcal{O}(KeySize)$, and line 33 is $\mathcal{O}(V + E)$. The total computational time complexity considering the outer loops ($EL$) and upon simplification is: $\mathcal{O}(E^2L^2)$. $L$ will typically be a small value making the time complexity $\mathcal{O}(E^2)$.



\section{Results \& Analysis}\label{sec:results} 
In this section, we analyze the effectiveness of the proposed framework (X-DFS) in creating DFS strategies to counter reverse engineering attacks (SAIL \cite{chakraborty2021sail, Chakraborty2019SAIL:}, OMLA \cite{alrahis2021omla}, SWEEP \cite{alaql2019sweep}). We do not consider the SAT \cite{Subramanyan2015Evaluating} attack for this study because: (1) It can be countered with well-known techniques such as Anti-SAT \cite{liu2020strong}, Full-Lock \cite{kamali2019full}, and LoPher \cite{saha2020lopher}; (2) It does not perform well for large designs and large key sizes due to its reliance on solving an NP-Hard problem in the backend; (3) It requires an oracle or unlocked design making it less practical. 


\begin{table*}[]
\captionsetup{justification=centering}
\caption{Ada X-DFS results for different values of $M$. We observe significant run-time improvement. Algo.~\ref{algo:XAIL_Lock} parameters: $KL$ = 10\% of Design Size, $A=TRUE$, $RN=FALSE$, $U=FALSE$, $Th_g=0.9$, $LC = 3$.}
\label{adaXail}
\renewcommand{\arraystretch}{1}
\small\addtolength{\tabcolsep}{-2.78pt}
\begin{tabular}{|
>{\columncolor[HTML]{ECF4FF}}c |
>{\columncolor[HTML]{FFFFFF}}c 
>{\columncolor[HTML]{FFFFFF}}c |
>{\columncolor[HTML]{FFFFFF}}c 
>{\columncolor[HTML]{FFFFFF}}c |
>{\columncolor[HTML]{FFFFFF}}c 
>{\columncolor[HTML]{FFFFFF}}c |
>{\columncolor[HTML]{FFFFFF}}c 
>{\columncolor[HTML]{FFFFFF}}c |}
\hline
\cellcolor[HTML]{EFE1C4}\textbf{Designs} & \multicolumn{2}{c|}{\cellcolor[HTML]{EFE1C4}\textbf{Ada X-DFS [$M=3$]}}                                            & \multicolumn{2}{c|}{\cellcolor[HTML]{EFE1C4}\textbf{Ada X-DFS [$M=7$]}}                                      & \multicolumn{2}{c|}{\cellcolor[HTML]{EFE1C4}\textbf{Ada X-DFS [$M=11$]}}                                     & \multicolumn{2}{c|}{\cellcolor[HTML]{EFE1C4}\textbf{General Settings}}                                     \\ \hline

\cellcolor[HTML]{E2DDDD}\textbf{\#}      & \multicolumn{1}{c|}{\cellcolor[HTML]{E2DDDD}\textbf{SAIL Acc}} & \cellcolor[HTML]{E2DDDD}\textbf{Lock Time (Sec)} & \multicolumn{1}{c|}{\cellcolor[HTML]{E2DDDD}\textbf{SAIL Acc}} & \cellcolor[HTML]{E2DDDD}\textbf{Lock Time (Sec)} & \multicolumn{1}{c|}{\cellcolor[HTML]{E2DDDD}\textbf{SAIL Acc}} & \cellcolor[HTML]{E2DDDD}\textbf{Lock Time (Sec)} & \multicolumn{1}{c|}{\cellcolor[HTML]{E2DDDD}\textbf{SAIL Acc}} & \cellcolor[HTML]{E2DDDD}\textbf{Lock Time (Sec)} \\ \hline

\textbf{sqrt}                            & \multicolumn{1}{c|}{\cellcolor[HTML]{FFFFFF}29.33}              &  289.78                                           & \multicolumn{1}{c|}{\cellcolor[HTML]{FFFFFF}17.44}              &   424.72                                        & \multicolumn{1}{c|}{\cellcolor[HTML]{FFFFFF}5.40}              &    575.19                                        & \multicolumn{1}{c|}{\cellcolor[HTML]{FFFFFF}2.10}                 &  1310.51                                          \\ \hline

\textbf{sin}                             & \multicolumn{1}{c|}{\cellcolor[HTML]{FFFFFF}14.33}                 & 43.62                                           & \multicolumn{1}{c|}{\cellcolor[HTML]{FFFFFF}7.96}                 &    73.92                                         & \multicolumn{1}{c|}{\cellcolor[HTML]{FFFFFF}4.02}                 &    105.03                                         & \multicolumn{1}{c|}{\cellcolor[HTML]{FFFFFF}1.47}                 &     194.34                                          \\ \hline

\textbf{div}                             & \multicolumn{1}{c|}{\cellcolor[HTML]{FFFFFF}26.65}                 &  520.41                                         & \multicolumn{1}{c|}{\cellcolor[HTML]{FFFFFF}19.04}                 & 818.16                                           & \multicolumn{1}{c|}{\cellcolor[HTML]{FFFFFF}19.99}                 &     1120.37                                        & \multicolumn{1}{c|}{\cellcolor[HTML]{FFFFFF}20.14}                 &  1678.42                                          \\ \hline

\textbf{arbiter}                         & \multicolumn{1}{c|}{\cellcolor[HTML]{FFFFFF}23.71}                 & 104.50                                           & \multicolumn{1}{c|}{\cellcolor[HTML]{FFFFFF}0.24}                 &    181.71                                          & \multicolumn{1}{c|}{\cellcolor[HTML]{FFFFFF}0.24}                 &      183.35                                       & \multicolumn{1}{c|}{\cellcolor[HTML]{FFFFFF}0.24 }                 &  183.35                                           \\ \hline

\textbf{memctrl}                         & \multicolumn{1}{c|}{\cellcolor[HTML]{FFFFFF}30.20}                 &  655.21                                        & \multicolumn{1}{c|}{\cellcolor[HTML]{FFFFFF}15.34}                & 985.47                                           & \multicolumn{1}{c|}{\cellcolor[HTML]{FFFFFF}15.18}              &   1355.06                                        & \multicolumn{1}{c|}{\cellcolor[HTML]{FFFFFF}13.03}                 & 1620.57                                            \\ \hline

\textbf{log2}                            & \multicolumn{1}{c|}{\cellcolor[HTML]{FFFFFF}28.93}                 &  366.53                                         & \multicolumn{1}{c|}{\cellcolor[HTML]{FFFFFF}12.84}                 &   884.01                                        & \multicolumn{1}{c|}{\cellcolor[HTML]{FFFFFF}11.10}                 &   1184.59                                         & \multicolumn{1}{c|}{\cellcolor[HTML]{FFFFFF}0.74}                 &  1784.23                                          \\ \hline

\textbf{square}                          & \multicolumn{1}{c|}{\cellcolor[HTML]{FFFFFF}28.10}                 &    366.38                                       & \multicolumn{1}{c|}{\cellcolor[HTML]{FFFFFF}22.43}                 &  592                                          & \multicolumn{1}{c|}{\cellcolor[HTML]{FFFFFF}13.43}                 &     855.71                                        & \multicolumn{1}{c|}{\cellcolor[HTML]{FFFFFF}13.56}                 &  1083.20                                         \\ \hline

\textbf{multiplier}                      & \multicolumn{1}{c|}{\cellcolor[HTML]{FFFFFF}27.27}              &  571.11                                         & \multicolumn{1}{c|}{\cellcolor[HTML]{FFFFFF}11.94}             &   807.53                                        & \multicolumn{1}{c|}{\cellcolor[HTML]{FFFFFF}23.76}             &       1156.57                                     & \multicolumn{1}{c|}{\cellcolor[HTML]{FFFFFF}0.69}             &  1695.14                                          \\ \hline

\textbf{voter}                           & \multicolumn{1}{c|}{\cellcolor[HTML]{FFFFFF}49.32}                 &  170.23                                          & \multicolumn{1}{c|}{\cellcolor[HTML]{FFFFFF}38.47}                 & 314.22                                            & \multicolumn{1}{c|}{\cellcolor[HTML]{FFFFFF}25.49}                 &   451.90                                        & \multicolumn{1}{c|}{\cellcolor[HTML]{FFFFFF}24.77}                 & 583.12                                           \\ \hline

\textbf{Average}                         & \multicolumn{1}{c|}{\cellcolor[HTML]{FFFFFF}\textbf{28.64}}     & \textbf{343.03}                                 & \multicolumn{1}{c|}{\cellcolor[HTML]{FFFFFF}\textbf{16.19}}     & \textbf{564.63}                                  & \multicolumn{1}{c|}{\cellcolor[HTML]{FFFFFF}\textbf{13.17}}     & \textbf{776.42}                                  & \multicolumn{1}{c|}{\cellcolor[HTML]{FFFFFF}\textbf{8.52}}     & \textbf{1128.06}                                  \\ \hline

\end{tabular}%
\end{table*}

\begin{table}[]
\captionsetup{justification=centering}
\caption{RN X-DFS can produce diverse (low similarity) SAIL-resilient locked designs. Algo.~\ref{algo:XAIL_Lock} parameters: $KL$ = 10\% of Design Size, $A=FALSE$, $RN=TRUE$, $U=TRUE$, $Th_g=0.9$, $LC = 3$.}
\label{RNXAIL}
\renewcommand{\arraystretch}{1}
\small\addtolength{\tabcolsep}{-3.5pt}
\begin{tabular}{|
>{\columncolor[HTML]{ECF4FF}}c |
>{\columncolor[HTML]{FFFFFF}}c 
>{\columncolor[HTML]{FFFFFF}}c |
>{\columncolor[HTML]{FFFFFF}}c 
>{\columncolor[HTML]{FFFFFF}}c |
>{\columncolor[HTML]{FFFFFF}}c 
>{\columncolor[HTML]{FFFFFF}}c |
>{\columncolor[HTML]{FFFFFF}}c 
>{\columncolor[HTML]{FFFFFF}}c |}
\hline
\cellcolor[HTML]{EFE1C4}\textbf{Designs} & \multicolumn{2}{c|}{\cellcolor[HTML]{EFE1C4}\textbf{Round 1}}                                                       & \multicolumn{2}{c|}{\cellcolor[HTML]{EFE1C4}\textbf{Round 2}}                                                       & \multicolumn{2}{c|}{\cellcolor[HTML]{EFE1C4}\textbf{Round 3}}                                                       & \multicolumn{2}{c|}{\cellcolor[HTML]{EFE1C4}\textbf{Round 4}}                                                       \\ \hline
\cellcolor[HTML]{E2DDDD}\textbf{\#}      & \multicolumn{1}{c|}{\cellcolor[HTML]{E2DDDD}\textbf{SAIL}} & \cellcolor[HTML]{E2DDDD}\textbf{Sim} & \multicolumn{1}{c|}{\cellcolor[HTML]{E2DDDD}\textbf{SAIL}} & \cellcolor[HTML]{E2DDDD}\textbf{Sim} & \multicolumn{1}{c|}{\cellcolor[HTML]{E2DDDD}\textbf{SAIL}} & \cellcolor[HTML]{E2DDDD}\textbf{Sim} & \multicolumn{1}{c|}{\cellcolor[HTML]{E2DDDD}\textbf{SAIL}} & \cellcolor[HTML]{E2DDDD}\textbf{Sim} \\ \hline
\textbf{sqrt}                            & \multicolumn{1}{c|}{\cellcolor[HTML]{FFFFFF}6.47}              & 0.132                                                & \multicolumn{1}{c|}{\cellcolor[HTML]{FFFFFF}7.78}              & 0.130                                                  & \multicolumn{1}{c|}{\cellcolor[HTML]{FFFFFF}7.62}              & 0.136                                                 & \multicolumn{1}{c|}{\cellcolor[HTML]{FFFFFF}8.11}                 & 0.132                                                  \\ \hline
\textbf{sin}                             & \multicolumn{1}{c|}{\cellcolor[HTML]{FFFFFF}7.11}              & 0.151                                              & \multicolumn{1}{c|}{\cellcolor[HTML]{FFFFFF}6.23}              & 0.157                                              & \multicolumn{1}{c|}{\cellcolor[HTML]{FFFFFF}6.71}              & 0.150                                             & \multicolumn{1}{c|}{\cellcolor[HTML]{FFFFFF}6.48}              & 0.140                                              \\ \hline
\textbf{div}                             & \multicolumn{1}{c|}{\cellcolor[HTML]{FFFFFF}7.91}                 & 0.228                                                  & \multicolumn{1}{c|}{\cellcolor[HTML]{FFFFFF}8.7}                 & 0.221                                                  & \multicolumn{1}{c|}{\cellcolor[HTML]{FFFFFF}10.3}                 & 0.230                                                  & \multicolumn{1}{c|}{\cellcolor[HTML]{FFFFFF}9.39}                 & 0.226                                                  \\ \hline
\textbf{arbiter}                         & \multicolumn{1}{c|}{\cellcolor[HTML]{FFFFFF}14.1}              & 0.410                                               & \multicolumn{1}{c|}{\cellcolor[HTML]{FFFFFF}12.93}              & 0.423                                               & \multicolumn{1}{c|}{\cellcolor[HTML]{FFFFFF}13.58}              & 0.473                                             & \multicolumn{1}{c|}{\cellcolor[HTML]{FFFFFF}14.06}              & 0.435                                              \\ \hline
\textbf{memctrl}                         & \multicolumn{1}{c|}{\cellcolor[HTML]{FFFFFF}10.8}              & 0.251                                                  & \multicolumn{1}{c|}{\cellcolor[HTML]{FFFFFF}10.50}                 & 0.265                                                  & \multicolumn{1}{c|}{\cellcolor[HTML]{FFFFFF}11.35}              & 0.241                                                  & \multicolumn{1}{c|}{\cellcolor[HTML]{FFFFFF}11.86}              & 0.260                                                  \\ \hline
\textbf{log2}                            & \multicolumn{1}{c|}{\cellcolor[HTML]{FFFFFF}8.84}              & 0.153                                                  & \multicolumn{1}{c|}{\cellcolor[HTML]{FFFFFF}10.70}              & 0.171                                                 & \multicolumn{1}{c|}{\cellcolor[HTML]{FFFFFF}9.55}              & 0.167                                                  & \multicolumn{1}{c|}{\cellcolor[HTML]{FFFFFF}9.43}              & 0.161                                                  \\ \hline
\textbf{square}                          & \multicolumn{1}{c|}{\cellcolor[HTML]{FFFFFF}9.18}                 & 0.210                                               & \multicolumn{1}{c|}{\cellcolor[HTML]{FFFFFF}8.22}              & 0.220                                              & \multicolumn{1}{c|}{\cellcolor[HTML]{FFFFFF}8.81}                 & 0.217                                                  & \multicolumn{1}{c|}{\cellcolor[HTML]{FFFFFF}8.70}                 & 0.227                                              \\ \hline
\textbf{multiplier}                      & \multicolumn{1}{c|}{\cellcolor[HTML]{FFFFFF}12.43}              & 0.193                                                  & \multicolumn{1}{c|}{\cellcolor[HTML]{FFFFFF}12.48}              & 0.194                                                  & \multicolumn{1}{c|}{\cellcolor[HTML]{FFFFFF}13.65}              & 0.192                                                  & \multicolumn{1}{c|}{\cellcolor[HTML]{FFFFFF}13.59}             & 0.192                                                  \\ \hline
\textbf{voter}                           & \multicolumn{1}{c|}{\cellcolor[HTML]{FFFFFF}23.30}              & 0.198                                              & \multicolumn{1}{c|}{\cellcolor[HTML]{FFFFFF}24.05}              & 0.186                                              & \multicolumn{1}{c|}{\cellcolor[HTML]{FFFFFF}20.93}              & 0.184                                             & \multicolumn{1}{c|}{\cellcolor[HTML]{FFFFFF}20.70}              & 0.230                                              \\ \hline
\textbf{Average}                         & \multicolumn{1}{c|}{\cellcolor[HTML]{FFFFFF}\textbf{11.12}}     & \textbf{0.214}                                     & \multicolumn{1}{c|}{\cellcolor[HTML]{FFFFFF}\textbf{11.34}}     & \textbf{0.218}                                     & \multicolumn{1}{c|}{\cellcolor[HTML]{FFFFFF}\textbf{11.39}}     & \textbf{0.221}                                     & \multicolumn{1}{c|}{\cellcolor[HTML]{FFFFFF}\textbf{11.37}}     & \textbf{0.225}                                     \\ \hline
\end{tabular}%
\end{table}

\subsection{Experimental Setup}
We implement the above described algorithms towards creating a highly parameterized framework for extensive evaluations. Algo.~\ref{algo:knowledge_extraction} was used to extract training data from a set of small designs from the ISCAS-85 benchmark suit (c1355, c1908, c2670, c3540, c5315, c6288, c7552 \cite{designs}) to speed up the learning process. However, during the evaluation process, we utilize large scale designs (sqrt, sin, div, arbiter, memctrl, log2, square, multiplier, voter \cite{designs}) following a transfer learning approach to better understand the scalability of the proposed framework. The locking dictionary ($lockDict$) has the following logic locking gate constructs: $XOR$, $XNOR$, $OR$, $AND$. We choose SAIL, SWEEP, and OMLA for our evaluation because: (1) They represent the large set of structural analysis-based attacks; (2) These attacks are cutting edge and strongest framework in structural attacks domain; (3) Each of them is highly scalable for large designs; and (4) Other functional attacks such as SAT attack can be mitigated using locking constructs such as Anti-SAT \cite{liu2020strong}, Full-Lock \cite{kamali2019full}, LoPher \cite{saha2020lopher}.

The SAIL and SWEEP attack models are trained on the following designs: c1355, c1908, c2670, c3540, c5315, c6288, c7552, sqrt, sin, div, arbiter, voter \cite{designs}.

On the contrary, we train the OMLA attack using the designs presented in \cite{OMLA}. We have located 1,000 designs for each of c1355, c1908, c2670, and c3540, totaling 4,000 designs. All designs have been utilized to train OMLA. The total number of gates defines the size of a design.

$keySize$ for Algo~\ref{algo:knowledge_extraction} are always set to $128$. $Th_{it}$ for Algo.~\ref{algo:knowledge_extraction} is set to 80. $KL$ for Algo.~\ref{algo:XAIL_Lock} has been altered in various ways for different experimental types, and we have specified all those key sizes in the subsequent descriptions.  Algo.~\ref{algo:knowledge_extraction} creates the dataset in just 1.5 hours with this settings. All experiments are run on: Intel Core i9-11900, 8 cores and 16 threads, 32GB RAM.



\subsection{X-DFS for Defending Against SAIL}
As shown in Table~\ref{XAIL_AI_Models}, we experiment utilizing different AI techniques for creating the X-DFS model to counter the SAIL attack. Among them, Adaboost outperforms all.
$10\%$ of each design size is the total amount (large quantity) of locking gates that X-DFS uses to lock the designs for this table. For instance, if a design has 30,000 gates, then 3,000 locking gates are inserted (10\%). Each value in this Table~\ref{XAIL_AI_Models} indicates SAIL's accuracy. This SAIL accuracy is the appropriate proportion of keys that the SAIL attack model identifies correctly. It appears that the div and the voter designs are inherently more vulnerable to SAIL, probably due to uniform structural patterns (similar to c6288 from ISCAS-85 benchmark suit as demonstrated in the original SAIL work \cite{chakraborty2021sail}). Table~\ref{XAIL_AI_Models} compares the effectiveness of different AI models for mitigating the SAIL attack using the X-DFS framework.


\begin{table*}[]
\centering
\captionsetup{justification=centering}
\caption{Ada X-DFS for varying thresholds ($Th_{g}$). Algo.~\ref{algo:XAIL_Lock} parameters: $KL$ = 10\% of Design Size, $A=TRUE$, $RN=FALSE$, $U=FALSE$, $M=11$, $LC = 3$.}
\label{table:th_result}
\renewcommand{\arraystretch}{1}
\small\addtolength{\tabcolsep}{5pt}
\begin{tabular}{|
>{\columncolor[HTML]{ECF4FF}}c |
>{\columncolor[HTML]{FFFFFF}}c 
>{\columncolor[HTML]{FFFFFF}}c |
>{\columncolor[HTML]{FFFFFF}}c 
>{\columncolor[HTML]{FFFFFF}}c |
>{\columncolor[HTML]{FFFFFF}}c 
>{\columncolor[HTML]{FFFFFF}}c |}
\hline
\cellcolor[HTML]{EFE1C4}\textbf{Designs} &
  \multicolumn{2}{c|}{\cellcolor[HTML]{EFE1C4}\textbf{Ada X-DFS [$Th_{g} > 0.50$]} } &
  \multicolumn{2}{c|}{\cellcolor[HTML]{EFE1C4}\textbf{Ada X-DFS [$Th_{g} > 0.65$]}} &
  \multicolumn{2}{c|}{\cellcolor[HTML]{EFE1C4}\textbf{Ada X-DFS [$Th_{g} > 0.80$]}} \\ \hline
\cellcolor[HTML]{E2DDDD}\textbf{\#} &
  \multicolumn{1}{c|}{\cellcolor[HTML]{E2DDDD}\textbf{SAIL Acc}} &
  \cellcolor[HTML]{E2DDDD}\textbf{Lock Time (Sec)} &
  \multicolumn{1}{c|}{\cellcolor[HTML]{E2DDDD}\textbf{SAIL Acc}} &
  \cellcolor[HTML]{E2DDDD}\textbf{Lock Time (Sec)} &
  \multicolumn{1}{c|}{\cellcolor[HTML]{E2DDDD}\textbf{SAIL Acc}} &
  \cellcolor[HTML]{E2DDDD}\textbf{Lock Time (Sec)}\\ \hline
\textbf{sqrt} &
  \multicolumn{1}{c|}{\cellcolor[HTML]{FFFFFF}13.98} &
  342.17 &
  \multicolumn{1}{c|}{\cellcolor[HTML]{FFFFFF}10.57} &
  367.39 &
  \multicolumn{1}{c|}{\cellcolor[HTML]{FFFFFF}7.18} &
  571.31 \\ \hline
\textbf{sin} &
  \multicolumn{1}{c|}{\cellcolor[HTML]{FFFFFF}18.63} &
  72.32 &
  \multicolumn{1}{c|}{\cellcolor[HTML]{FFFFFF}11.17} &
  83.41 &
  \multicolumn{1}{c|}{\cellcolor[HTML]{FFFFFF}6.67} &
 97.94 \\ \hline
\textbf{div} &
  \multicolumn{1}{c|}{\cellcolor[HTML]{FFFFFF}27.82} &
  998.07 &
  \multicolumn{1}{c|}{\cellcolor[HTML]{FFFFFF}18.69} &
  1035.44 &
  \multicolumn{1}{c|}{\cellcolor[HTML]{FFFFFF}18.05} &
  1117.22 \\ \hline
\textbf{arbiter} &
  \multicolumn{1}{c|}{\cellcolor[HTML]{FFFFFF}11.45} &
  67.33 &
  \multicolumn{1}{c|}{\cellcolor[HTML]{FFFFFF}9.21} &
  126.09 &
  \multicolumn{1}{c|}{\cellcolor[HTML]{FFFFFF}1.89} &
  155.07 \\ \hline
    \textbf{memctrl} &
  \multicolumn{1}{c|}{\cellcolor[HTML]{FFFFFF}18.32} &
  845.49 &
  \multicolumn{1}{c|}{\cellcolor[HTML]{FFFFFF}14.16} &
  920.13 &
  \multicolumn{1}{c|}{\cellcolor[HTML]{FFFFFF}15.91} &
  1239.39 \\ \hline
\textbf{log2} &
  \multicolumn{1}{c|}{\cellcolor[HTML]{FFFFFF}23.2} &
  823.11 &
  \multicolumn{1}{c|}{\cellcolor[HTML]{FFFFFF}17.4} &
  897.27 &
  \multicolumn{1}{c|}{\cellcolor[HTML]{FFFFFF}13.47} &
  1106.4 \\ \hline
\textbf{square} &
  \multicolumn{1}{c|}{\cellcolor[HTML]{FFFFFF}18.43} &
  619.54 &
  \multicolumn{1}{c|}{\cellcolor[HTML]{FFFFFF}19.11} &
  834.39 &
  \multicolumn{1}{c|}{\cellcolor[HTML]{FFFFFF}13.22} &
  843.15 \\ \hline
\textbf{multiplier} &
  \multicolumn{1}{c|}{\cellcolor[HTML]{FFFFFF}24.05} &
  778.29 &
  \multicolumn{1}{c|}{\cellcolor[HTML]{FFFFFF}18.29} &
  907.52 &
  \multicolumn{1}{c|}{\cellcolor[HTML]{FFFFFF}24.01} &
  1078.71 \\ \hline
\textbf{voter} &
  \multicolumn{1}{c|}{\cellcolor[HTML]{FFFFFF}26.59} &
  243.02 &
  \multicolumn{1}{c|}{\cellcolor[HTML]{FFFFFF}21.83} &
  330.02 &
  \multicolumn{1}{c|}{\cellcolor[HTML]{FFFFFF}25.11} &
  398.26 \\ \hline
\textbf{Average} &
  \multicolumn{1}{c|}{\cellcolor[HTML]{FFFFFF}\textbf{20.27}} &
  \textbf{532.15} &
  \multicolumn{1}{c|}{\cellcolor[HTML]{FFFFFF}\textbf{15.6}} &
  \textbf{611.3} &
  \multicolumn{1}{c|}{\cellcolor[HTML]{FFFFFF}\textbf{13.95}} &
  \textbf{734.16} \\ \hline
\end{tabular}%

\end{table*}


\subsection{Comparison Against Other LL Techniques}
In Table~\ref{otherLocking}, we compare X-DFS with other popular logic locking algorithms, including SFLL\_Point \cite{NEOS, yasin2019sfll}. Here, the random locking (column $Random$) uses the same set of locking dictionary as X-DFS without the added learning/intelligence. X-DFS clearly outperforms all these techniques in terms of building up SAIL resiliency. Also note that Anti-SAT fails to lock some of the designs (ran for $> 20$ hours) probably due to their large size and other algorithm specific constraints (indicated with $-$, inside the table). For the experiments in Table~\ref{otherLocking}, these big designs are locked using $128$ bit keys. This is because some logic locking methods are unable to lock designs with larger keys ( $> 128$ can not be handled). This is, however, not a limitation for X-DFS, which is highly scalable in this regard.

\subsection{X-DFS for Countering OMLA and SWEEP}
In Table~\ref{omla_Result} and Table~\ref{table:sweep_result} we illustrate the effectiveness of X-DFS against OMLA \cite{alrahis2021omla} and SWEEP \cite{alaql2019sweep} respectively. We utilize the transfer learning approach for these experiments by using the X-DFS model already trained for mitigating SAIL.
As seen in Table~\ref{omla_Result} and Table~\ref{table:sweep_result}, X-DFS shows great performance in terms of mitigating OMLA and SWEEP even when it is not trained on those specific data. This ability to provide protecting against unknown attacks is one of the most crucial aspects of building a logic locking (or DFS) framework, since new attacks will inevitably arise. X-DFS is quite effective in accomplishing this goal. This demonstrates the practicality and robustness of X-DFS. Results shown in Table~\ref{omla_Result} are provided for $64$ bit keys because the publicly available trained model and the dataset for OMLA \cite{OMLA} utilize $64$ bit keys. 
For SWEEP and OMLA, we observed the best performance when utilizing Random Forest (among other machine learning algorithms).
\begin{table}[]
\centering
\captionsetup{justification=centering}
\caption{X-DFS for different $LC$. Values indicate SAIL attack accuracy. Algo.~\ref{algo:XAIL_Lock} parameters: $KL$ = 10\% of Design Size, $A=FALSE$, $RN=FALSE$, $U=TRUE$.}
\label{table:locality_Result}
\renewcommand{\arraystretch}{1}
\small\addtolength{\tabcolsep}{1pt}
\begin{tabular}{|
>{\columncolor[HTML]{ECF4FF}}c |c|c|c|c|}
\hline
\cellcolor[HTML]{E2DDDD}\textbf{Designs} &
  \cellcolor[HTML]{E2DDDD}\textbf{Random} &
  \cellcolor[HTML]{E2DDDD}\textbf{$LC = 5$} &
  \cellcolor[HTML]{E2DDDD}\textbf{$LC = 7$} &
  \cellcolor[HTML]{E2DDDD}\textbf{$LC = 10$} \\ \hline
\textbf{sqrt}       & 55.89          & 6.89          & 5.46         & 3.94          \\ \hline
\textbf{sin}        & 53.9           & 5.62          & 4.43          & 2.5           \\ \hline
\textbf{div}       & 57.08          & 8.67          & 10.95         & 4.37        \\ \hline
\textbf{arbiter}    & 53.75          & 3.3           & 3.91          & 3.79          \\ \hline
\textbf{memctrl}    & 57.97          & 1.12          & 0.9           & 0.86          \\ \hline
\textbf{log2}       & 53.73          & 2.19          & 3.58          & 1.94          \\ \hline
\textbf{square}     & 53.17          & 4.81          & 2.12          & 4.4           \\ \hline
\textbf{multiplier} & 55.52          & 4.98          & 5.24          & 4.65          \\ \hline
\textbf{voter}      & 53.58          & 5.1           & 15.18         & 8.32          \\ \hline
\textbf{Average}    & \textbf{54.96} & \textbf{4.74} & \textbf{5.75} & \textbf{3.86} \\ \hline
\end{tabular}%

\end{table}
\subsection{Effectiveness of X-DFS for Different Locality Sizes}
In Table~\ref{table:locality_Result} we present the X-DFS results for countering SAIL attack while using structural features with locality sizes $5$, $7$, and $10$. X-DFS generally performs better as we increase the locality size. Analyzing a larger locality provides more context for the AI model in terms of determining the optimal locking choices. All the results in this Table~\ref{table:th_result} are obtained using the Adaboost model.


\subsection{Improving the Runtime: Ada X-DFS}
\begin{figure}[!t]
\centering

\includegraphics[width=\linewidth]{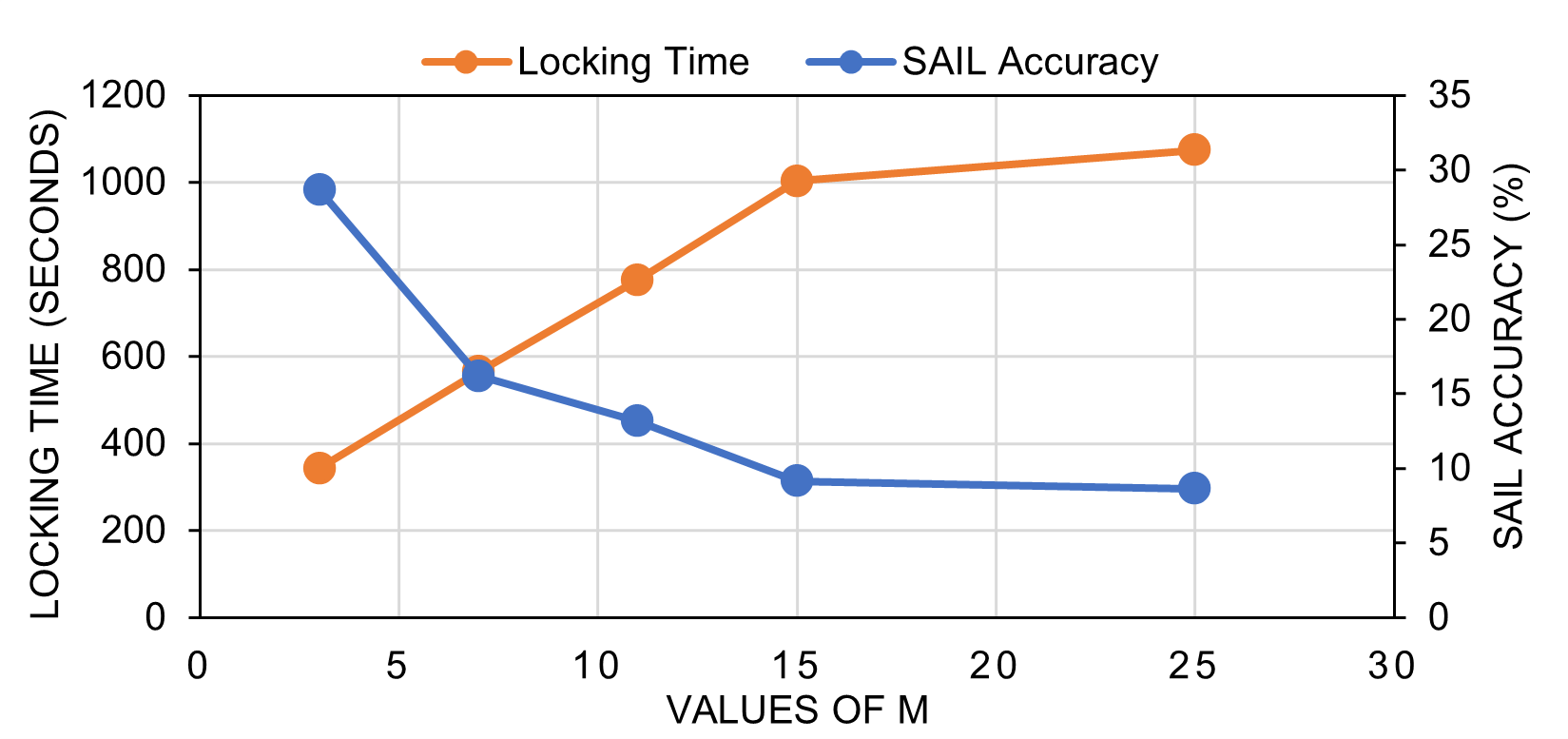}
\caption{Trade off between X-DFS speed and efficiency. Algo.~\ref{algo:XAIL_Lock} parameters same as Table~\ref{adaXail}. \label{fig:Performance_Tradeoff}}

\end{figure}

The X-DFS locking algorithm (Algo~\ref{algo:XAIL_Lock}) has a time complexity of $\mathcal{O}(E^2)$ which is already very scalable (analysis provided above). However, to further speed things up, we introduce an approximation using the adaptive flag ($A$) as shown in Algo~\ref{algo:XAIL_Lock}. We set $A = TRUE$, $RN = FALSE$, and vary $M = \{3, 7, 11\}$ (M selects the least number of possibilities, mentioned in Algo.~\ref{algo:XAIL_Lock}) toward obtaining the results in Table~\ref{adaXail}. For this table, $Good$ (in Algo~\ref{algo:XAIL_Lock}) is measured using $Th_{g}\ge0.90$. Since synchronization with other locking approaches is not required, larger size keys ($10\%$ of design size) are utilized to track how long X-DFS takes to produce resilient designs. It shows a clear trade-off between run-time and SAIL resilience in Fig.~\ref{fig:Performance_Tradeoff}. As $M$ increases, we observe that both the execution time and SAIL resilience improve. This is because higher values of $M$ result in a larger search space, allowing for the identification of better DFS/locking candidates. 
The \textit{General Settings} column of Table~\ref{adaXail} showcases the results of X-DFS when $A = FALSE$ (non Adaptive).

\subsection{Ada X-DFS with different thresholds}
Next we examine the behavior of the framework for different threshold ($Th_{g}$) values, with $M$ being held constant at $11$. These results are shown in Table~\ref{table:th_result}. When the threshold ($Th_{g}$) is increased to values above $0.50$, we observed that the locking duration increases and the accuracy of the SAIL attack decreases. This is because a higher threshold allows the X-DFS to search for more optimal locking candidates.

\subsection{Introducing Randomness: RN X-DFS}
X-DFS is an informed (through learning) locking technique, but some degree of randomness is necessary for different designing scenarios (for obfuscations or variant generations). Hence, we have allowed for some controlled randomness using the flag $RN$ in Algo~\ref{algo:XAIL_Lock}. We set $A = FALSE$, $RN = TRUE$, towards obtaining the results in Table~\ref{RNXAIL} for multiple instances of X-DFS locking. We also show through a cross-instance averaged cosine similarity metric ($Sim$ column in Table~\ref{RNXAIL}) in several rounds that the generated designs are indeed locked differently while maintaining a high degree of SAIL resilience ($SAIL$ column in Table~\ref{RNXAIL}). 



\begin{table}[]
\centering
\captionsetup{justification=centering}
\caption{Directly locking with extracted rules using LeGO. Values indicate SAIL attack accuracy. KeySize, $KL$ = 10\% of Design Size is used.}
\label{table:ruleLocking}
\renewcommand{\arraystretch}{1}
\small\addtolength{\tabcolsep}{7pt}
\begin{tabular}{|c|c|c|}
\hline
\rowcolor[HTML]{EFEDED} 
\multicolumn{1}{|c|}{\cellcolor[HTML]{E2DDDD}\textbf{Designs}}         & \multicolumn{1}{c|}{\cellcolor[HTML]{E2DDDD}\textbf{Random}} & \multicolumn{1}{c|}{\cellcolor[HTML]{E2DDDD}\textbf{X-DFS Rules + LeGO}}  
\\ \hline
\multicolumn{1}{|c|}{\cellcolor[HTML]{ECF4FF}\textbf{sqrt}}      & \multicolumn{1}{c|}{55.89}                                                                   & \multicolumn{1}{c|}{16.11}                                                                        
\\ \hline
\multicolumn{1}{|c|}{\cellcolor[HTML]{ECF4FF}\textbf{sin}}      & \multicolumn{1}{c|}{53.90}                                                                & \multicolumn{1}{c|}{14.89}                                                             
\\ \hline
\multicolumn{1}{|c|}{\cellcolor[HTML]{ECF4FF}\textbf{div}}      & \multicolumn{1}{c|}{57.08}                                                             & \multicolumn{1}{c|}{18.13}                                                                                   
\\ \hline
\multicolumn{1}{|c|}{\cellcolor[HTML]{ECF4FF}\textbf{arbiter}}     & \multicolumn{1}{c|}{53.75}                                                                 & \multicolumn{1}{c|}{11.06}                                                                                    
\\ \hline
\multicolumn{1}{|c|}{\cellcolor[HTML]{ECF4FF}\textbf{memctrl}}      & \multicolumn{1}{c|}{57.97}                                                                 & \multicolumn{1}{c|}{17.67}                                                                                        
\\ \hline
\multicolumn{1}{|c|}{\cellcolor[HTML]{ECF4FF}\textbf{log2}}      & \multicolumn{1}{c|}{53.73}                                                                    & \multicolumn{1}{c|}{19.90}                                                                                                                   
\\ \hline
\multicolumn{1}{|c|}{\cellcolor[HTML]{ECF4FF}\textbf{square}}      & \multicolumn{1}{c|}{53.17}                                                                 & \multicolumn{1}{c|}{20.01}                                                                                                             
\\ \hline
\multicolumn{1}{|c|}{\cellcolor[HTML]{ECF4FF}\textbf{multiplier}}      & \multicolumn{1}{c|}{55.52}                                       & \multicolumn{1}{c|}{15.62}                                                                                      
\\ \hline
\multicolumn{1}{|c|}{\cellcolor[HTML]{ECF4FF}\textbf{voter}}      & \multicolumn{1}{c|}{53.58}                                                             & \multicolumn{1}{c|}{29.29}                                                                              
\\ \hline

\multicolumn{1}{|c|}{\cellcolor[HTML]{ECF4FF}\textbf{Average}}                            & \multicolumn{1}{c|}{\textbf{54.96}}                                                              & \multicolumn{1}{c|}{\textbf{18.07}}                                            
\\ \hline

\end{tabular}
\end{table}

\begin{figure*}[!t]
\centering

\includegraphics[width=0.95\linewidth]{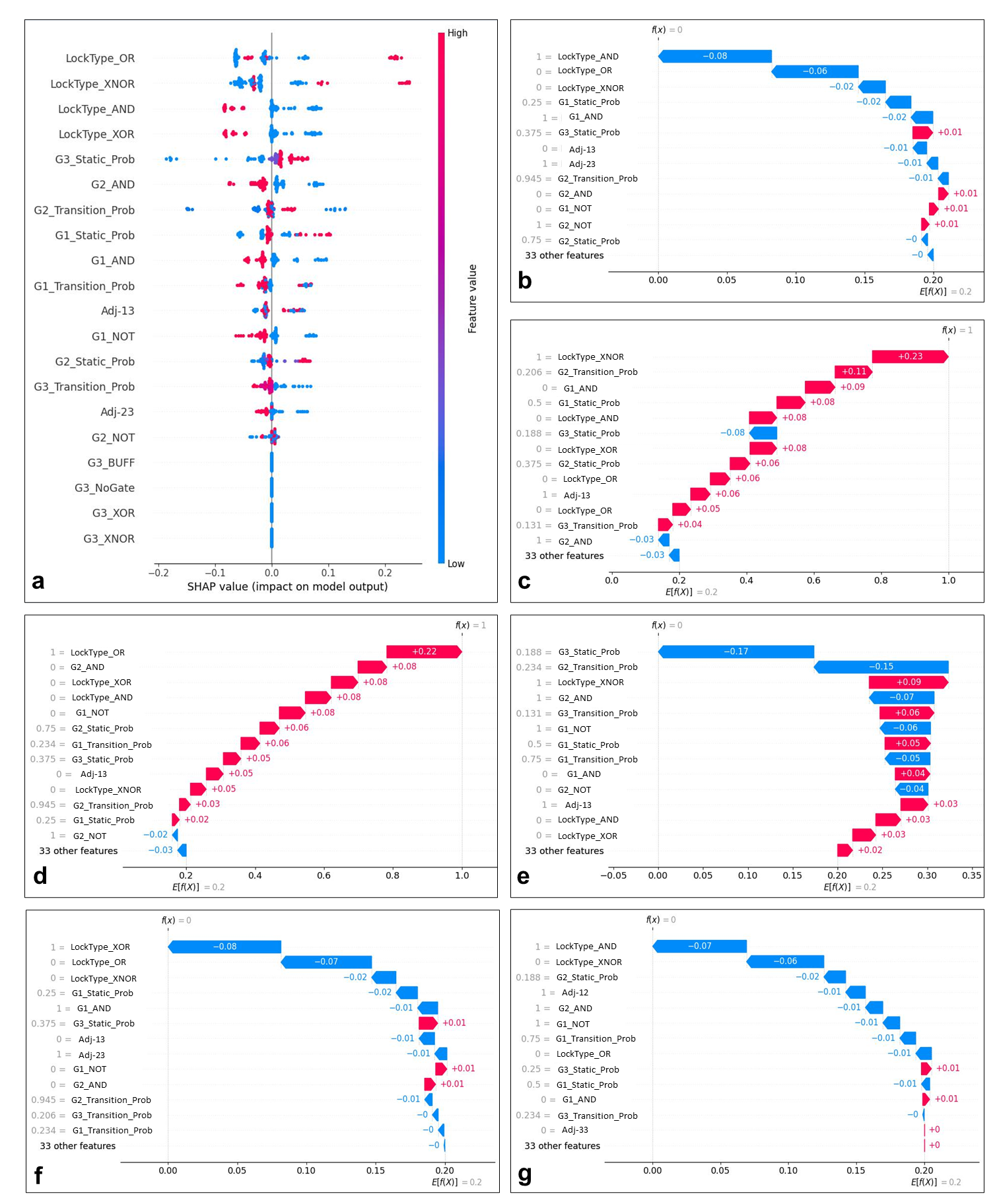}
\caption{Explainability of X-DFS\_RF Model. From top to bottom: (a) SHAP-based feature summary extracted from 100 random data points; (b-g) SHAP waterfall plots for different samples.\label{fig:SVM_XAI}}

\vspace{-0.2in}
\end{figure*}
\begin{figure*}[!t]
\centering

\includegraphics[width=0.95\linewidth]{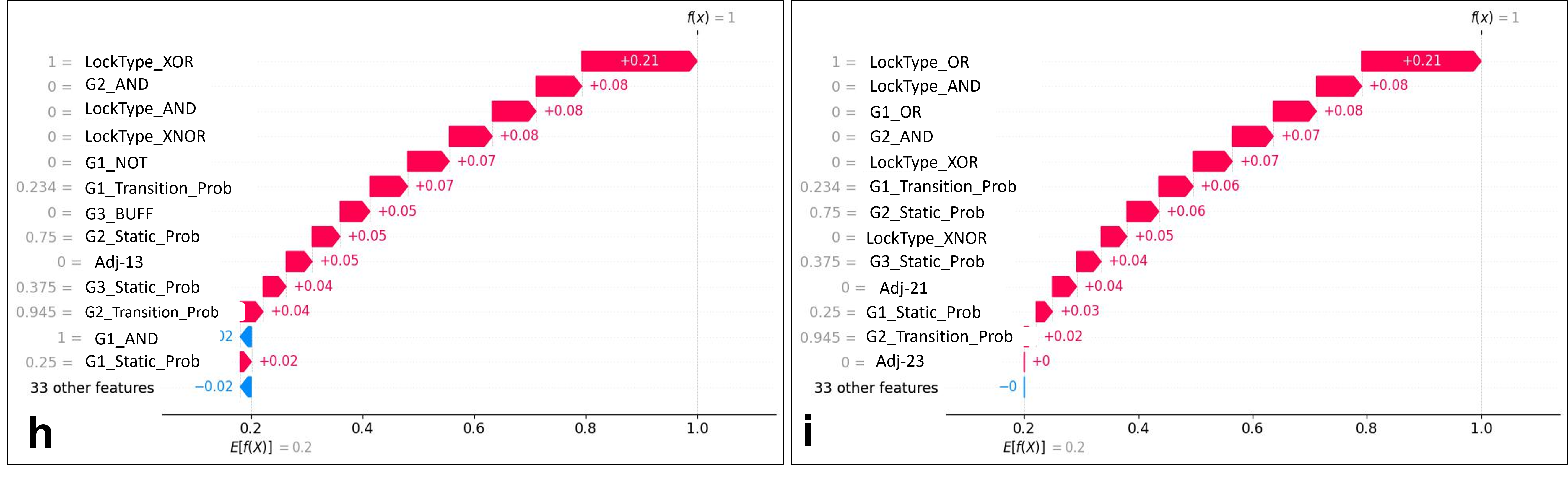}
\caption{(h-i) Explainability of X-DFS\_RF Model: Additional SHAP waterfall plots.\label{fig:add_rules}}

\end{figure*}
\begin{table*}[ht]
\caption{Human-readable rules from X-DFS: Logic Locking Case Study.}
\label{table:extractedRules}
\renewcommand{\arraystretch}{1}
\small\addtolength{\tabcolsep}{-4.3pt}
\begin{tabular}{|
>{\columncolor[HTML]{FFCCC9}}c |
>{\columncolor[HTML]{FFFFFF}}c |
>{\columncolor[HTML]{FFFFFF}}c |}
\hline
\cellcolor[HTML]{EFEFEF}\textbf{} &
  \cellcolor[HTML]{EFEFEF}\textbf{IF} &
  \cellcolor[HTML]{EFEFEF}\textbf{Action} \\ \hline
\cellcolor[HTML]{DAE8FC}\textbf{b} &
  G1 = AND \&\& Static probability of G1 = low &
  \cellcolor[HTML]{FFFFFF}Do not lock with AND/XOR \\ \hline
\textbf{c} &
  G1 != AND \&\& Transition probability of G2 = low \&\& Static probability of G1 = moderate &
  \cellcolor[HTML]{FFFFFF}Lock with XNOR \\ \hline
\textbf{d} &
  G2 != AND \&\& G1 != NOT \&\& Static Probability of G2 = high &
  \cellcolor[HTML]{FFFFFF}Lock with OR \\ \hline
\cellcolor[HTML]{DAE8FC}\textbf{e} &
  Transition probability of G2 = low \&\& G2 = AND \&\& G1 = NOT &
  Do not lock with any gate \\ \hline
\cellcolor[HTML]{DAE8FC}\textbf{f} &
  Static probability of G1 = low \&\& G1 = AND \&\& G2 and G3 connected &
  Do not lock with XOR \\ \hline
\cellcolor[HTML]{DAE8FC}\textbf{g} &
  Static probability of G2 is low \&\& G2 == AND \&\& a NOT gate is connected to an AND gate &
  Do not lock with AND \\ \hline
\textbf{h} &
  \cellcolor[HTML]{FFFFFF}\begin{tabular}[c]{@{}c@{}} G1 != NOT \&\& Static probability of G1 = high \&\& G2 != AND \&\&\\ Transition probability of G2 = low \&\& G3 != BUFF \&\& Static probability of G3 = moderate\end{tabular} &
  Lock with XOR \\ \hline
\textbf{i} &
  \cellcolor[HTML]{FFFFFF}\begin{tabular}[c]{@{}c@{}} Transition probability of G1 = low \&\& Static probability of G1 = low \&\&\\ Static probability of G2 = high \&\& Transition probability of G2 = high \&\& G1 != OR \&\& G2 != AND\end{tabular} &
  Lock with OR \\ \hline
\end{tabular}
\vspace{-0.2in}
\end{table*}

\subsection{Explainability: Which Features are Important for X-DFS?}
The top sub-figures in Fig.~\ref{fig:SVM_XAI}(a) shows the feature importance summary plot for the X-DFS\_RF\_Model (on 100 random data points). The feature names are mentioned on the Y-Axis ranked in descending order of significance. The position on the X-axis (with respect to the neural point 0.0) determines the amount of positive (right side, towards label 1) or negative (left side, towards label 0) impact. Each dot is a data point which represents a row of data from the original dataset. Each point on the graph is assigned a color that corresponds to the value of the related feature. High values are represented by the color red, while low values are represented by the color blue. $LockType$ refers to a specific type of locking gate. G1 (Gate1), G2 (Gate2), and G3 (Gate3) are the three gates in the structural feature locality (in Fig.~\ref{fig:structFeat}) and the value next to them indicates their gate type. The naming of these gates are based on their breadth-first-search order as detailed in Section~\ref{sec:features}. Static\_Prob is referring to static probability (Table~\ref{table:static_feat}) and Transition\_Prob is referring to transition probability (Table~\ref{transproba}). We can observe that locking with gate types $OR$ and $XNOR$ is preferred by the model, whereas locking with gate types $AND$ and $XOR$ is avoided. Selecting a wire with higher static probability for locking is preferred. It is also evident that design regions with $AND$ gates are avoided during locking. SHAP \cite{SHAP} utilizes a game-theory-based approach and some of the features do not participate strongly enough (dropped from these plots).





\subsection{Explainability: Can we Extract Good Locking Rules?}
Using SHAP algorithms, it is also possible to understand the decision process for each sample by observing the waterfall graphs (b,c,d,e,f,g in Fig.~\ref{fig:SVM_XAI}). The waterfall plot represents the values of the X-axis corresponding to the target variable, which in this case is the probability of good locking. x represents the selected observation, f(x) is the predicted value of the model for input x, and E[f(x)] represents the expected value of the target variable, which is essentially the average of all predictions (in our case, 100 random points). The SHAP value for each feature is indicated by the length of the bar (effect on prediction). For example, in Fig.~\ref{fig:SVM_XAI}(c) locking type $XNOR$ is pushing the prediction towards 1. 

From these plots, we can extract `rules' that are being implicitly used by the X-DFS\_Models for locking a given design to counter SAIL, OMLA, and SWEEP attacks. We consider probability values below 0.3 as low, 0.3 to 0.5 as moderate, and beyond 0.5 as high. Rules are shown in Table~\ref{table:extractedRules}. These rules (extracted via X-DFS) can also be used to directly lock designs using manual methods or frameworks like LeGO \cite{alaql2021lego}. This knowledge can be used to understand the strengths and weaknesses of a given attack vector and thereby assist researchers/engineers in mitigating similar attacks/variants.

\subsection{Directly Locking with Extracted Rules: LeGO}
The human-understandable rules generated from X-DFS can be directly utilized by the LeGO \cite{alaql2021lego} framework to lock a design. In Table~\ref{table:ruleLocking}, we show the effectiveness of using the X-DFS rules (Table~\ref{table:extractedRules}) directly with the help of the LeGO framework for mitigating SAIL attack.


\section{Conclusion}
\label{sec:conclusion}
Through this work, we have developed an explainable artificial intelligence guided framework (X-DFS) that can automatically navigate the design-for-security search space for generating a set of mitigation rules for a given novel attack vector. We have explained the inner-workings of X-DFS using detailed algorithms and implemented the framework towards creating a highly parameterized tool (specifically for reverse engineering attacks). Using the implantation, we have demonstrated the effectiveness of the framework for mitigating logic locking attacks such as SAIL, SWEEP, and OMLA in diverse settings. To boost the speed of X-DFS, we introduce X-DFS\_Ada which can operate on large designs with up to 60,000 gates within 15 minutes and still ensure attack resiliency. Another variant, X-DFS\_RN, was also introduced for injecting controlled randomness in the process towards diverting certain type of bias-driven attacks. X-DFS has outperformed most existing popular logic locking DFS algorithms including SFLL\_Point. We have also demonstrated how X-DFS can be used to understand the decision process of AI models towards creating human-understandable X-DFS rules for countering target attacks. 
Future works will focus on exploring the effectiveness of X-DFS against other attack vectors (e.g., hardware Trojans, fault injection, side channel) and developing more sophisticated explainable algorithms.

\section{Acknowledgment}
This material is based upon work supported by the National Science Foundation (NSF) under Grant No. 2350363 and Grant No. 2316399.





\bibliographystyle{IEEEtran}
\bibliography{IEEEabrv,XAIL}

\end{document}